\documentclass[preprint]{elsarticle}

\usepackage{bm,amsmath,amsfonts,amssymb}
\usepackage{hyperref}       
\usepackage{graphicx} 
\usepackage{nicefrac}       
\usepackage{url}            
\usepackage{doi}
\usepackage{orcidlink}
\usepackage{verbatim}
\usepackage{xcolor}

\begin{document}

\begin{frontmatter}
\title{A Machine Learning Based Approach to Online Electron Reconstruction at CLAS12}

\author[1]{R.~Tyson\,\orcidlink{0000-0002-0635-4198}}
\author[1]{G.~Gavalian\,\orcidlink{0000-0002-6738-5457}}
\affiliation[1]{organization={Thomas Jefferson National Accelerator Facility}, 
                city={Newport News}, 
                postcode={VA 23606},
                country={USA}}

\begin{abstract}
Online reconstruction is key for monitoring purposes and real time analysis in High Energy and Nuclear Physics experiments. A necessary component of reconstruction algorithms is particle identification that combines information left by a particle passing through several detector components to identify the particle’s type. Of particular interest to electro-production Nuclear Physics experiments such as CLAS12 is electron identification which is used to trigger data recording. A machine learning approach was developed for CLAS12 to reconstruct and identify electrons by combining raw signals at the data acquisition level from several detector components. This approach achieves an electron identification purity above 75\% whilst retaining an efficiency close to 100\%. The machine learning tools are capable of running at high rates exceeding the data acquisition rates and will allow electron reconstruction in real-time. This work enhances online analyses and monitoring and can contribute to improved triggering at CLAS12. This machine learning driven approach will also be crucial for experiments aiming to transition to streaming readout operations where online reconstruction will be a key component of the data taking paradigm.

\end{abstract}

\end{frontmatter}

\section{Introduction} \label{sect_intro}
\noindent
High Energy and Nuclear Physics experiments are producing increasing large quantities of data, that need to be efficiently processed in real-time towards data preservation. Typically this relies on incremental trigger levels that filter the data based on hardware and physics requirements. Online monitoring and real time analysis are also crucial elements of High Energy and Nuclear Physics experimental operations as they ensure that the experiment delivers good quality data. Online reconstruction and particle identification schemes are key to triggering, online data selection and real time analysis. These schemes aim to process raw signals at the data acquisition level from several detector components in order to determine what original particles passed through a particle detector and their characteristics. \\

\noindent
The use of machine learning (ML) for online reconstruction is attractive due to high processing rates capable of keeping up with the data acquisition rate. Recent work explored the use of Convolutional Neural Networks (CNNs)~\cite{AITrigger,CNN1} for electron triggering at the CEBAF Large Acceptance Spectrometer at 12 GeV (CLAS12) in order to improve the electron trigger purity~\cite{AITrigger}. Refs.~\cite{OnlineElID1,OnlineElID2} showed how electron identification could be improved in the ATLAS trigger using neural networks, allowing to halve background levels. Machine learning algorithms such as boosted decision trees are also employed in the High Level Triggers at the LHCb~\cite{LHCbTrig}. In Ref.~\cite{OnlineVertex}, neural networks were used to accurately reconstruct the vertex of a track from micropattern gaseous detector data. Ref.\cite{MuonTrig} demonstrates how muon identification for triggering at the Compact Muon Solenoid (CMS) could be improved by using a neural network to remove noise from the Drift Tubes and a second network to identify the side of passage of the hit with respect to a Drift Tube wire. In this article we will describe how online electron reconstruction can be achieved for CLAS12. Going one step further than previous approaches, the procedure proposed here will not only identify the presence of an electron but fully reconstruct all necessary information for real time analysis and physics reaction identification.\\

\noindent
The Continuous Electron Beam Facility (CEBAF) is located at the Thomas Jefferson National Accelerator Facility (JLab)~\cite{CEBAF2,JLAB1}. CEBAF delivers an electron beam with energies up to 11 GeV to all four experimental halls at JLab~\cite{JLAB1}, with plans to increase the beam energy up to 12 GeV over the coming years. The CLAS12 detector is located in Hall B and aims to further the global understanding of hadronic structure and Quantum Chromodynamics~\cite{C12Overview}. A diagram of CLAS12 is shown in Figure~\ref{fig:FD}. The CLAS12 detector has almost complete azimuthal angle coverage, with low polar angular coverage (2.5-5$^o$) enabled by the Forward Tagger, forward angular coverage (5-35$^o$) enabled by the Forward Detector and large polar angular coverage enabled by the Central Detector (35-125$^o$)~\cite{C12Overview,C12Det}. The Forward Detector is further segmented into six sectors in azimuthal angle. Electrons with very low scattering angles are detected in the Forward Tagger, by using the coincidence of a track in the Forward Tagger tracker and a hit in the Forward Tagger hodoscope which allows to differentiate between electrons and photons~\cite{C12Det}. Due to low backgrounds and simple detector apparatus, electron detection and identification in the Forward Tagger is rather straightforward. Electrons are not identified in the Central Detector. In the Forward Detector (FD), electron reconstruction is a more involved process, and will be the focus of the discussion presented in this article.\\

\noindent
Online reconstruction at CLAS12 would be beneficial for improved online monitoring, real time analysis and improved triggering. In 2023, the first steps towards full online reconstruction were taken, where tracks from the FD drift chambers were reconstructed in the online operation of CLAS12 experiments~\cite{InstaREC}. This was notably beneficial as CLAS12 then ran experiments with dual targets, and the tracking allowed to reconstruct the electron vertex which allowed to monitor the production of electrons from both targets. This online track reconstruction was based on ML tracking efforts at CLAS12 that are already incorporated in the offline reconstruction and which yielded increases in tracking efficiency of up to 50\% for certain channels, with a decrease in computing time of roughly 30\%~\cite{AI_Tracking,AI_Tracking2,denoise}. Building on these AI predicted tracks, subsequent neural networks were able to predict the momentum and angles of the tracks with good resolution~\cite{TrackParam}. This avoids costly track fitting procedures, allowing to produce reconstructed tracks during online data taking operations. However, the tracks were not identified as belonging to a given particle type, meaning that they cannot be used for more advanced analysis or triggering. As mentioned previously, a proof of principle application of CNNs to the electron trigger was found to be a promising avenue for improved triggering~\cite{AITrigger}. However the drawback here is that the implementation suggested in Ref.~\cite{AITrigger} would only identify the presence of an electron, without assigning it to a track and fully reconstructing it. Additionally, CNNs are relatively slow compared to simpler neural networks.\\

\noindent
This article will present a proof of concept algorithm for full online electron reconstruction at CLAS12. Simple and fast Multilayer Perceptrons (MLP)~\cite{MLP} type algorithms will be combined to the online track reconstruction algorithm to identify electrons from the data acquisition level information of CLAS12 subsystems used for electron identification. Section~\ref{sect_ml} will give a brief description of the machine learning algorithms used in this article. Section~\ref{sect_conv} will present a more in depth overview of offline electron reconstruction at CLAS12. Section~\ref{sect_algo} will describe how tracks can be associated to hits in subsequent detector subsystems and how this information can be used to identify electrons. Section~\ref{sect_tests} will describe further tests demonstrating the good performance of these algorithms for online electron reconstruction. Section~\ref{sect_retrain} will then describe improvements to the training sample that allow the online electron identification to achieve electron identification efficiencies close to 100\%. Section~\ref{sect_ccl} will then give brief conclusions and outlook.

\section{Brief Overview of Deep Learning} \label{sect_ml}
\noindent
Deep learning algorithms such as neural networks are popular for machine learning applications due to their versatility, good performance and high prediction rates. Many popular packages and libraries exist to facilitate the implementation of deep learning algorithms. As the CLAS12 reconstruction framework is based in Java~\cite{C12Software}, the Java based DEEP NETTS~\cite{DEEPNETTS} library was used in this article.\\

\noindent
Multilayer perceptrons are feed-forward neural networks consisting of interconnected nodes, arranged in layers, with each node in a layer connected to nodes in subsequent layers~\cite{MLP}. The output of a node is given by:

\begin{equation*}
    o_k = f\left(\sum_i w_{ik}o_i\right),
\end{equation*}

\noindent
where $w_{ik}$ denotes the weight for node $i$ to $k$ and $f$ denotes an activation function, such as the ReLu function $f(x) = max(0,x)$. In a binary classification tasks where the aim of the neural network is to output the probability that an event belongs to one of two classes, the output of the final layer is passed through a sigmoid function. In regression tasks where the aim of the network is to predict one or more quantities, the output of the final layer is typically passed through a linear activation function. Neural networks are trained using variations of Stochastic Gradient Descent algorithms such as ADAM~\cite{adam}. Backpropagation algorithms modify the weights of each node given gradients so as to minimise a loss function, such as the cross entropy loss function for classification tasks and the mean squared error for regression tasks.\\

\noindent
Different neural network types are best suited to certain applications. Convolutional neural networks~\cite{CNN1} are well suited for image based problems. They are composed of convolutional layers to enhance or remove features of inputted images, and a feed-forward neural network that processes the output of the convolutional layers. Auto-encoders are well suited to generative tasks. They are composed of two networks, an encoder which transforms an input into a latent space representation which is then processed by a decoder which has an output typically of the same dimensions as the input to the auto-encoder. In this article, the networks used towards online electron reconstruction are mostly multilayer perceptrons. The reason for this choice is that these networks are comparatively less complicated than other types of neural networks, leading to higher prediction rates.\\

\section{Offline Electron Reconstruction in the CLAS12 Forward Detector} \label{sect_conv}

\begin{figure}[ht!]
    \centering   
    \includegraphics[width=0.99\textwidth]{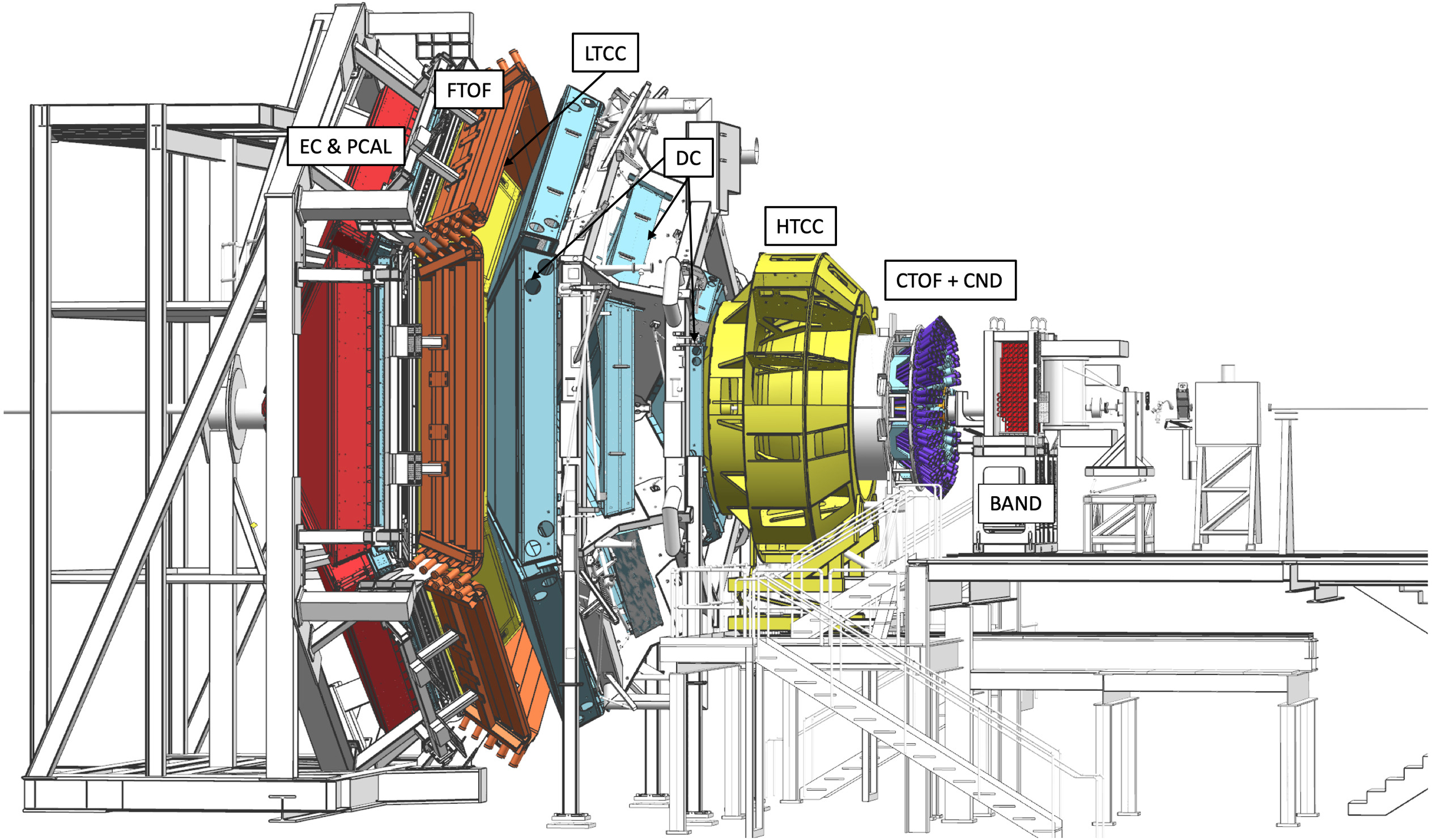}
    \caption[The CLAS12 Forward Detector.]{The CLAS12 Detector, with highlighted detector subsystems. Image taken from~\cite{C12Overview}.}
    \label{fig:FD}
\end{figure}

\noindent
Electron reconstruction in the CLAS12 Forward Detector (FD) involves several different detectors, shown in Figure~\ref{fig:FD}. The FD Drift Chambers (DC) allow to reconstruct the momentum and angles of charged particles and in particular electrons~\cite{C12DC}. These tracks and their associated three momentum can then be associated to hits in other CLAS12 detectors~\cite{C12Software}.  The CLAS12 High Threshold Cherenkov Counters (HTCC) acts as a veto to identify electrons by requiring at least two photoelectrons produced in the HTCC~\cite{C12HTCC}. Electrons will deposit large amounts of energy in the CLAS12 Forward Electromagnetic Calorimeters (ECAL)~\cite{C12ECAL}. This allows to distinguish electrons from other particle types such as pions that will deposit lower amounts of energy in the calorimeters. \\

\begin{figure}[ht!]
    \centering   
    \includegraphics[width=0.8\textwidth]{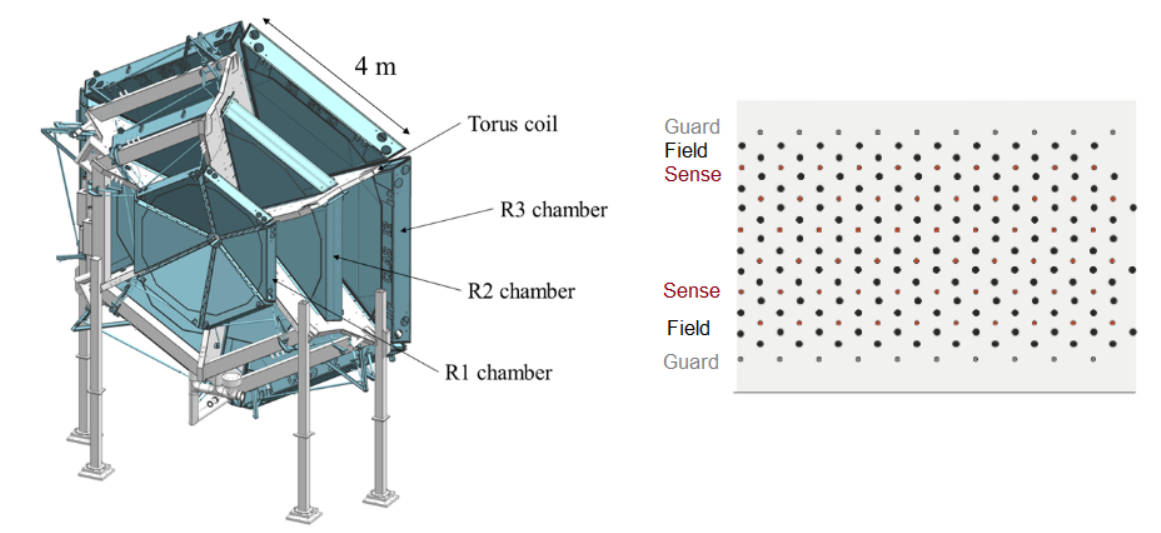}
    \caption[The CLAS12 Drift Chambers.]{A schematic of the CLAS12 FD drift chambers (left) and one of its superlayers (right). Images taken from~\cite{C12DC}}
    \label{fig:DC_Both}
\end{figure}

\noindent
The DC is composed of six superlayers organised in three regions in each of the FD's six sectors. Each superlayer is composed of six layers, with 112 sense wires per layer~\cite{C12DC}. A schematic of the DC and of one of its superlayers is shown in Figure~\ref{fig:DC_Both}. The trajectory of a particle going through the DC is recorded as the particle ionises gas contained in the DC. The electrons released from the gas are then detected by the sense wires closest to the particle's path. The charge and momentum of the particle are measured based on the curvature of the particle's path in a magnetic field with known strength, with the second region of the DC immersed in the magnetic field. The reconstruction process first creates clusters from hits in adjacent wires within a layer of the DC. Segments of clusters are then formed over a superlayer from clusters in different layers of the superlayer. Fits to the intersection of segments between superlayers then allow to form tracks in the drift chambers and obtain the track parameters such as momentum and angles. The track fitting employs a Kalman Filter, which also returns a $\chi^2$ assessment of the track quality. \\

\noindent
Machine learning algorithms were found to significantly increase the FD tracking efficiency and decrease the time spent on tracking. A denoising stage implemented with a convolutional auto-encoder decreases the amount of noise in the DC~\cite{denoise}. A multilayer perceptron is used to find good positive and negative track candidates from the average wire position of each superlayer segments~\cite{AI_Tracking,AI_Tracking2}. A track corruption auto encoder allows to recover missing superlayer segments, which then allows to recover tracks due to a requirement to have at least 5 superlayer segments~\cite{AI_Tracking,AI_Tracking2}.  Removing noise allows for a better average wire position to be assigned to a cluster. This allows for more realistic track segments and in turn tracks with lower $\chi^2$ that might otherwise be discarded, as does the track suggestion algorithm that finds good tracks. Additionally, the denoising and track suggestion algorithms lead to fewer fake tracks which in turn decreases the number of real tracks lost due to overlapping track segments. Overall, the use of machine learning in tracking at CLAS12 lead to a 50\% gain in tracking efficiency. The total tracking time is also decreased by roughly 30\% as there are fewer tracks to fit.\\

\begin{figure}[ht!]
    \centering   
    \includegraphics[width=0.8\textwidth]{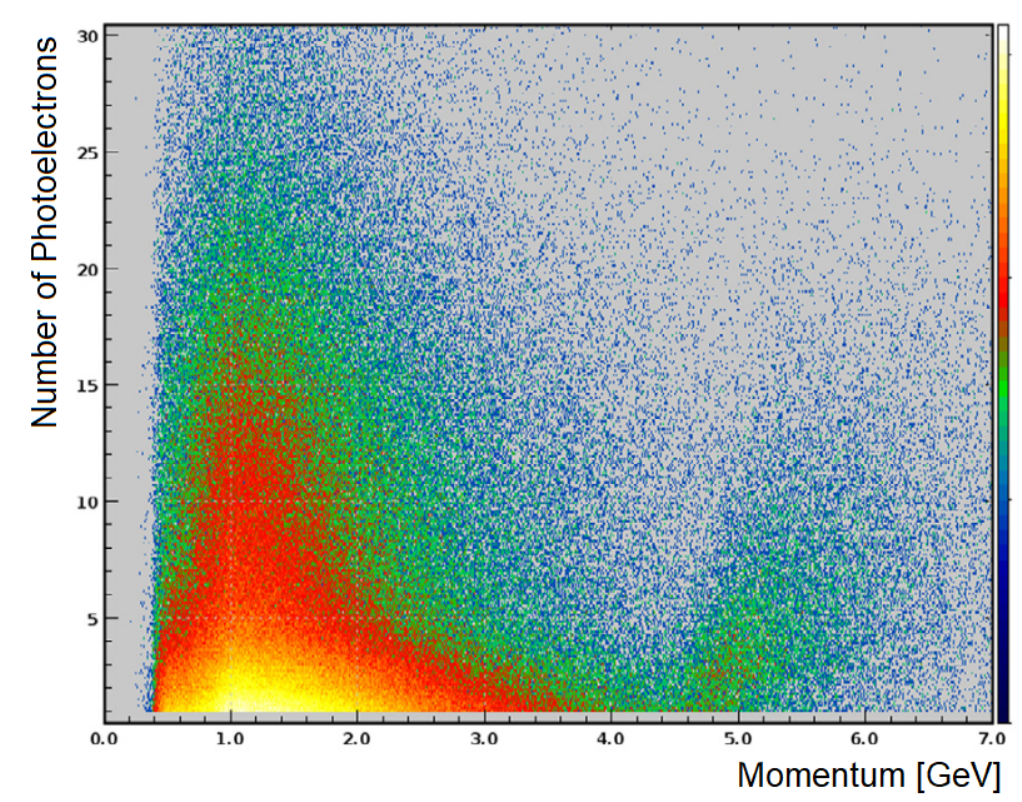}
    \caption[The number of photo-electrons recorded in the HTCC as a function of particle momentum.]{The number of photo-electrons recorded in the HTCC as a function of momentum for positively charged particles. Images taken from~\cite{C12DC}}
    \label{fig:HTCC}
\end{figure}

\noindent
The HTCC was built to identify electrons and positrons and separate them from other charged particle types~\cite{C12HTCC}. The HTCC is composed of eight mirrors in each of the six FD sectors. The HTCC mirrors will produce Cherenkov radiation when an electron travels through the mirror, with the radiated photons read out by photo-multiplier tubes (PMTs). Clusters are formed from PMTs collecting light from adjacent mirrors. The number of photo-electrons associated with a particle then becomes a discriminating variable in electron identification. However, pions will start to produce more than two photoelectrons when they have momentum greater than 4.5 to 5 GeV. This effect is clearly demonstrated in Figure~\ref{fig:HTCC} which shows an increase above 5 GeV in reconstructed clusters associated to positively charged particles, leading to positive pions being identified as positrons.\\

\begin{figure}[ht!]
\centering  
\includegraphics[width=0.49\textwidth]{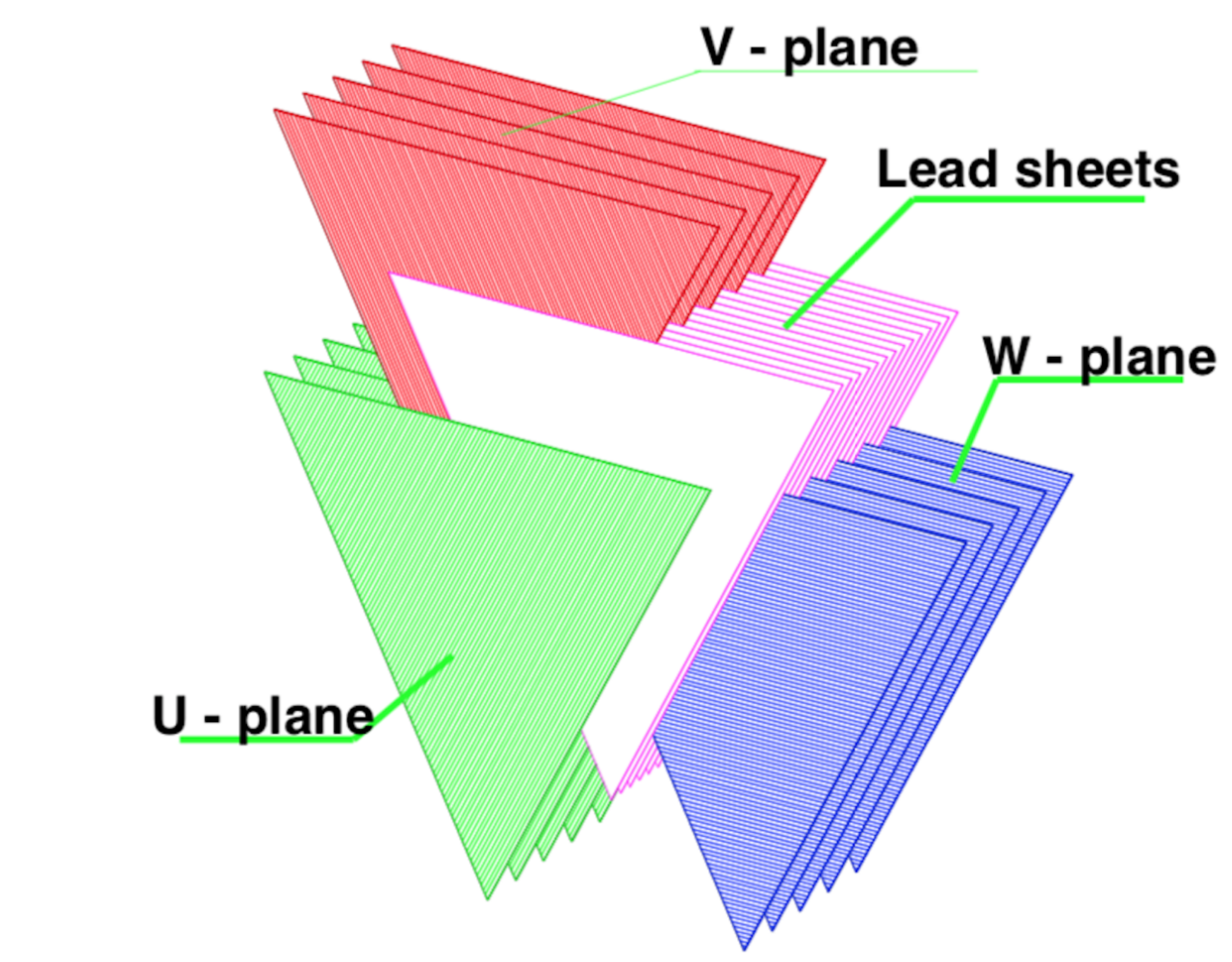}
\includegraphics[width=0.49\textwidth]{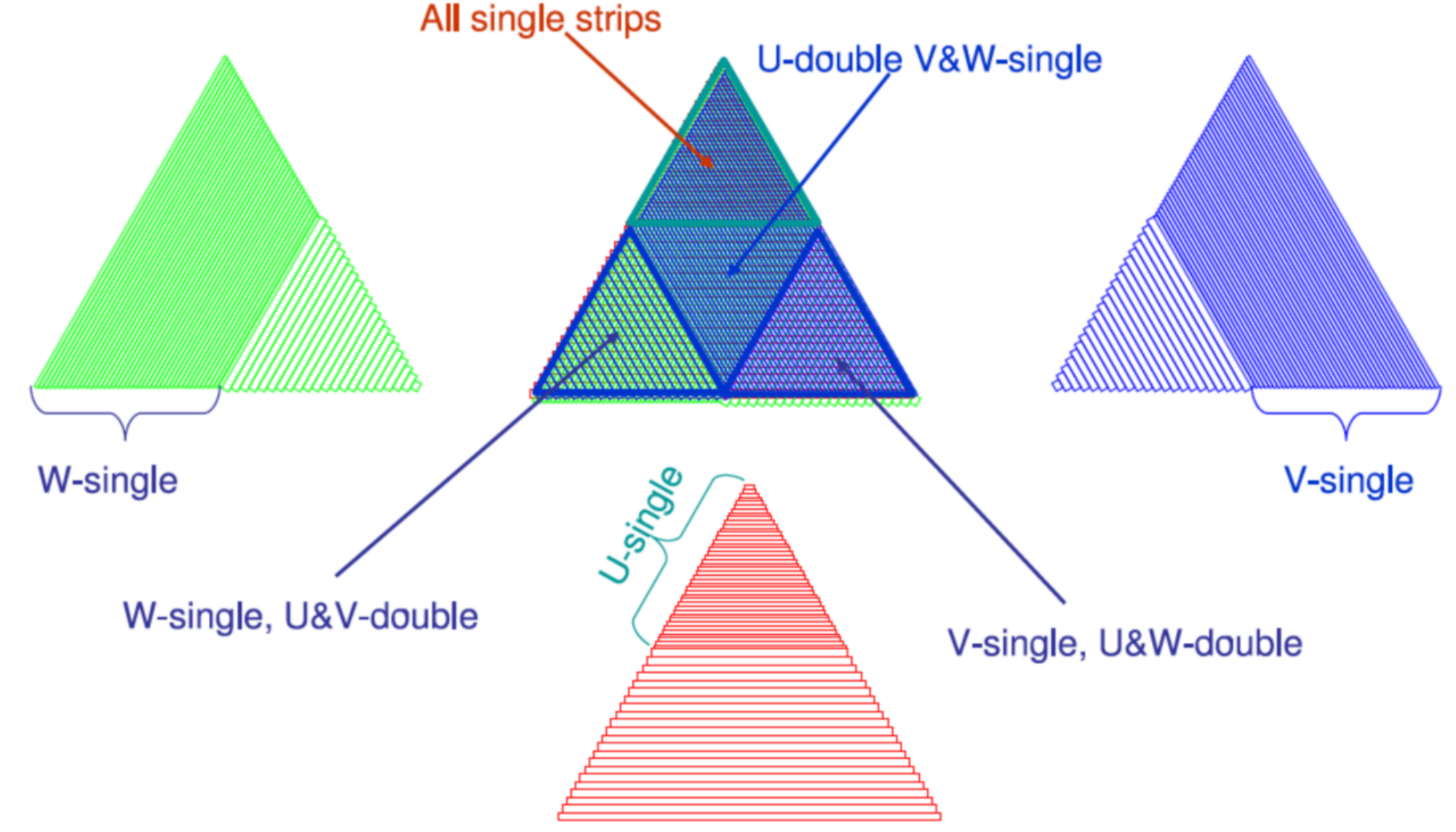}
\caption[Schematic representation of the calorimeters' three views.]{Schematic representation of the calorimeters' three views. Image taken from~\cite{C12ECAL}.}
\label{fig:ECALDiag}
\end{figure}

\noindent
The ECAL is composed of three layers, a preshower calorimeter (PCAL) and inner and outer layers (ECIN and ECOUT respectively). All three layers have three views (U/V/W), with the PCAL having 68 strips in U and 62 in V and W, whilst the ECIN and ECOUT have 36 strips in all three views~\cite{C12ECAL}. Figure~\ref{fig:ECALDiag} shows a schematic representation of the ECAL's three views. Particles will deposit energy in the calorimeter as they travel through it. A hit is defined as a strip having energy above a certain threshold. Peaks are defined from adjacent hits in a view, with clusters identified at the intersection of a peak in each view. The position and time of clusters are used to calculate the velocity of neutral particles, allowing to distinguish between photons and neutrons. Photons and electrons will produce electromagnetic showers with high energy deposition. A cluster which is correlated in space with a DC track can then be identified as an electron based on its energy deposition.\\

\noindent
The Forward Time of Flight Detector (FTOF) was designed to measure the velocity of charged hadrons so as to assign a particle type to these hadrons~\cite{C12FTOF}. The FTOF is composed of three arrays of scintillator counters, named panel-1a, panel-1b, panel-2. Panel-1b has the better timing resolution as it contains 62 counters instead of 23 in panel-1a and 5 in panel-2. PMTs are used to detect the scintillation light produced by a particle travelling through the FTOF. Clusters are formed from adjacent PMTs producing a signal above a certain threshold. The cluster time is associated to the known position of the scintillation counter to calculate the velocity of the particle passing through the FTOF. Particle types are uniquely identified by their charge and mass. The mass of a particle can be related to its velocity and momentum as $m=\frac{p}{\beta\gamma}$, for $\gamma$ the Lorentz factor and $\beta=\frac{v}{c}$ with $c$ the speed of light. The FTOF can then be used in conjunction with the DC to identify charged particles. This relies on the vertex time at which the event started (known as the event start time). This can be calculated from electrons that can be assumed to have a $\beta=1$ at CLAS12 energies and resolutions, allowing to calculate the start time given that $\beta=\frac{d}{(t_{FTOF}-t_{st})c}$ for $t_{FTOF}$ the FTOF time and $t_{st}$ the event start time. Electron identification is therefore a necessary first step to identifying other particle types at CLAS12.\\

\begin{figure}[ht!]
    \centering   
    \includegraphics[width=0.8\textwidth]{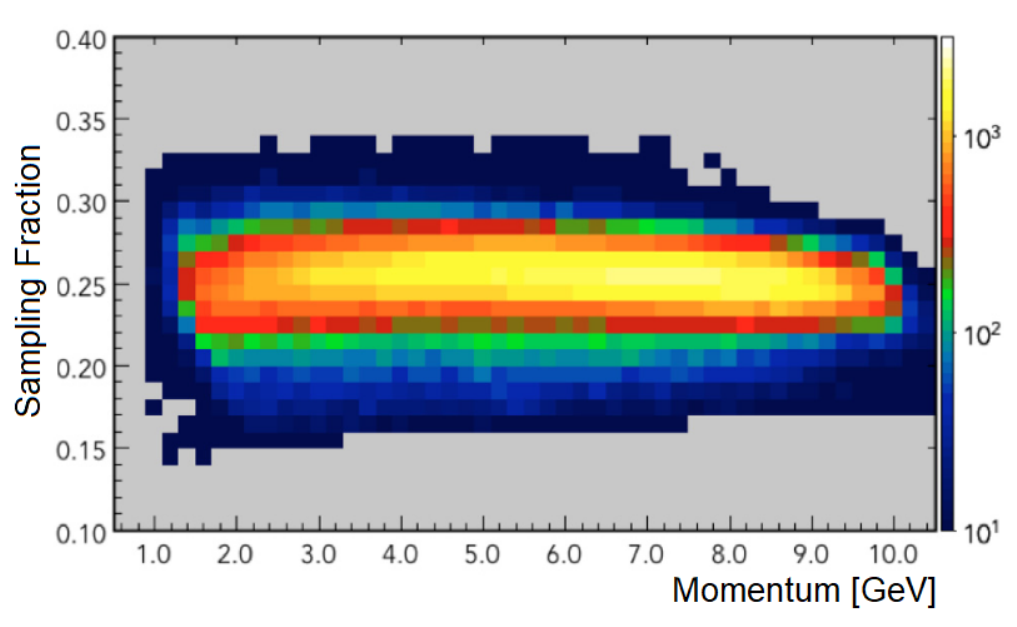}
    \caption[The electron sampling fraction.]{The electron sampling fraction as a function of momentum. Images taken from~\cite{C12ECAL}}
    \label{fig:SF}
\end{figure}

\noindent
The reconstruction algorithm combines information from each of the DC, HTCC and ECAL to identify electrons~\cite{C12Software}. The CLAS12 swimmer algorithm will use the measured CLAS12 magnetic field to geometrically match a track with given vertex and momentum to hits in the HTCC and ECAL~\cite{C12Software}. Electrons are then required to produce at least two photoelectrons in the HTCC. They will also be required to deposit at least 60 MeV in the PCAL. The sampling fraction, defined as the sum of the energy deposition in each of the PCAL, ECIN and ECOUT divided by the particle momentum should be constant for electrons around 0.25. The electron sampling fraction as a function of momentum is shown in Figure~\ref{fig:SF}. The sampling fraction for electrons is fitted with a Gaussian in slices of momentum. Electrons are then required to be within 5$\sigma$ of the mean, with both the $\sigma$ and mean parameterised as a function of momentum. Charged hadrons are identified with the time of flight technique based on their velocity measured by the CLAS12 Forward Time of Flight (FTOF) detector~\cite{C12Software} as described previously.\\

\noindent
The trigger system at CLAS12 uses electron identification in the FD as it main trigger~\cite{C12Trigger}. The requirements placed on electron identification are similar to those in the reconstruction. However, one drawback is that there only is a loose geometrical matching between the DC tracks and hits in other detectors and no calculation of the sampling fraction as the momentum of a track will not be known at the trigger level. The trigger is then susceptible to increased particle misidentification background compared to the reconstruction. The trigger can be run with or without DC roads to improve geometrical matching. The CLAS12 trigger system also allows for other measurements such as photoproduction with low angle electron scattering by requiring an electron in the Forward Tagger and two tracks in the FD. However, the trigger system has some limitations such as not being able to identify two tracks in the same sector. Finally, a J/$\psi$ trigger was added to some CLAS12 run periods that required two minimum ionising particle tracks with low energy deposition in the ECAL and in opposite sectors of the FD.\\

\section{Online Electron Reconstruction in the CLAS12 Forward Detector} \label{sect_algo}

\noindent
Several of the conventional algorithms used for offline electron reconstruction in the FD at CLAS12 are too slow to be used online. Ref.~\cite{InstaREC} details how the machine learning track finding and track corruption algorithms can be implemented to produce tracks in the online reconstruction. The denoising algorithm is too slow to be used online, but the same track efficiency can be achieved by removing the requirement to avoid overlapping tracks. Fitting tracks with a Kalman Filter is a computationally intensive task and is too slow for the online reconstruction. This is instead replaced with a multilayer perceptron trained to predict the track momentum and direction, as described in Refs.~\cite{InstaREC,TrackParam}. The combinations of these two algorithms are currently deployed online at CLAS12, allowing to produce tracks and reconstruct their momentum and angles during data taking operations at CLAS12. However, these tracks are not identified as belonging to specific particle types. Here we describe how electrons can be identified given the tracks reconstructed online.\\

\noindent
In the offline reconstruction, the next step after reconstructing tracks is to associate these to clusters in the HTCC and the ECAL. This utilises the CLAS12 swimmer which is again too slow to be used in the online reconstruction. The first step of the online electron reconstruction algorithm is therefore to replace the CLAS12 swimmer with a faster algorithm. A multilayer perceptron can be employed to this end. The input to the network is the average wire position of the reconstructed tracks in each DC superlayer. This is the same input as what is given to the online track reconstruction algorithm~\cite{TrackParam} and contains enough information to infer the trajectory of the track in CLAS12. The output is the particle's cluster peak strip number on the front face of each of the PCAL, ECIN and ECOUT layers, in all three U/V/W views. The network is also trained to predict the FTOF panel-1B cluster position and path to the FTOF panel-1B. The network has four hidden layers with a tanh activation function and 12, 36, 27 and 18 nodes respectively, whilst the output layer has a linear activation function.\\

\begin{figure}[ht!]
    \centering   
    \includegraphics[width=0.44\textwidth]{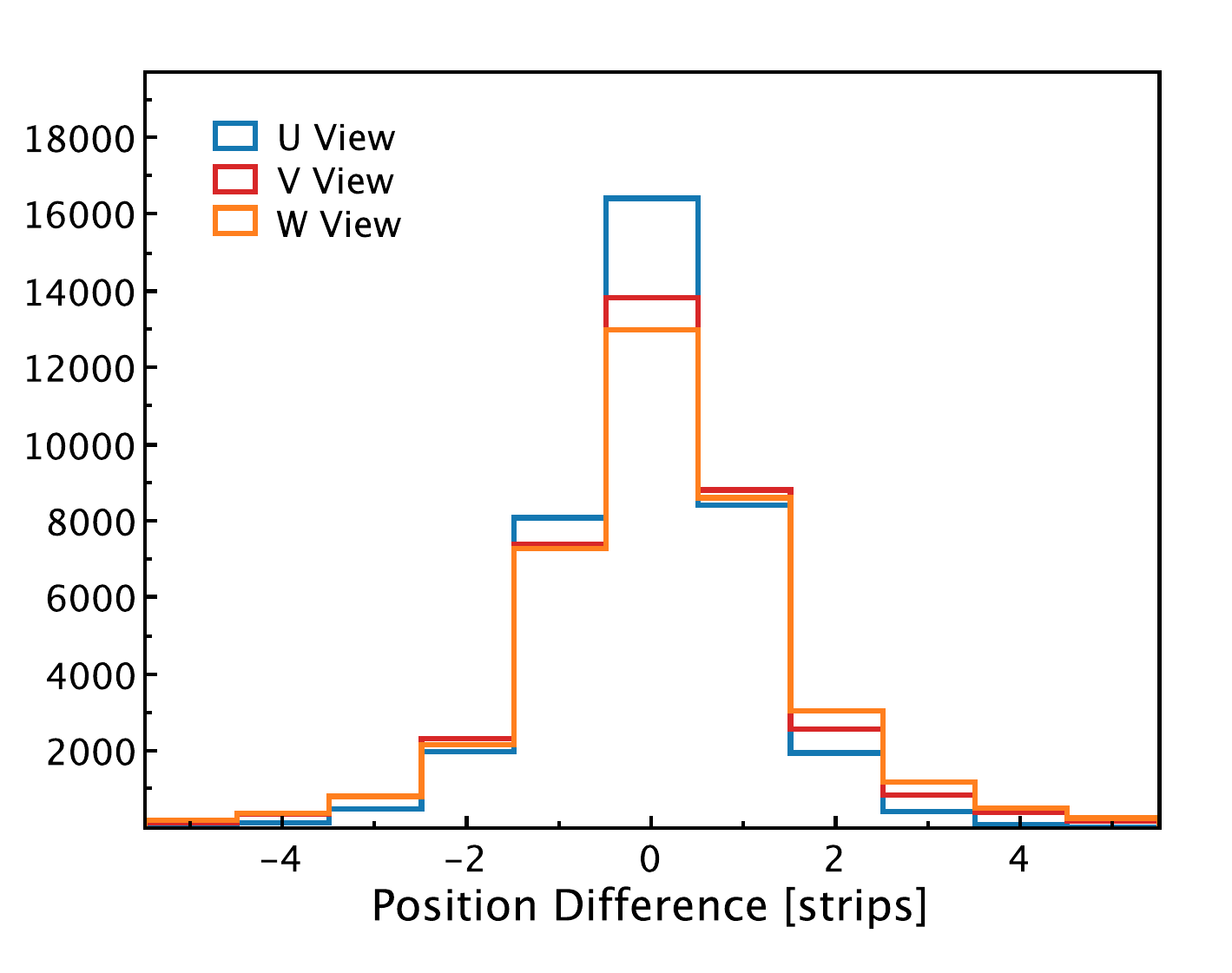}
    \includegraphics[width=0.44\textwidth]{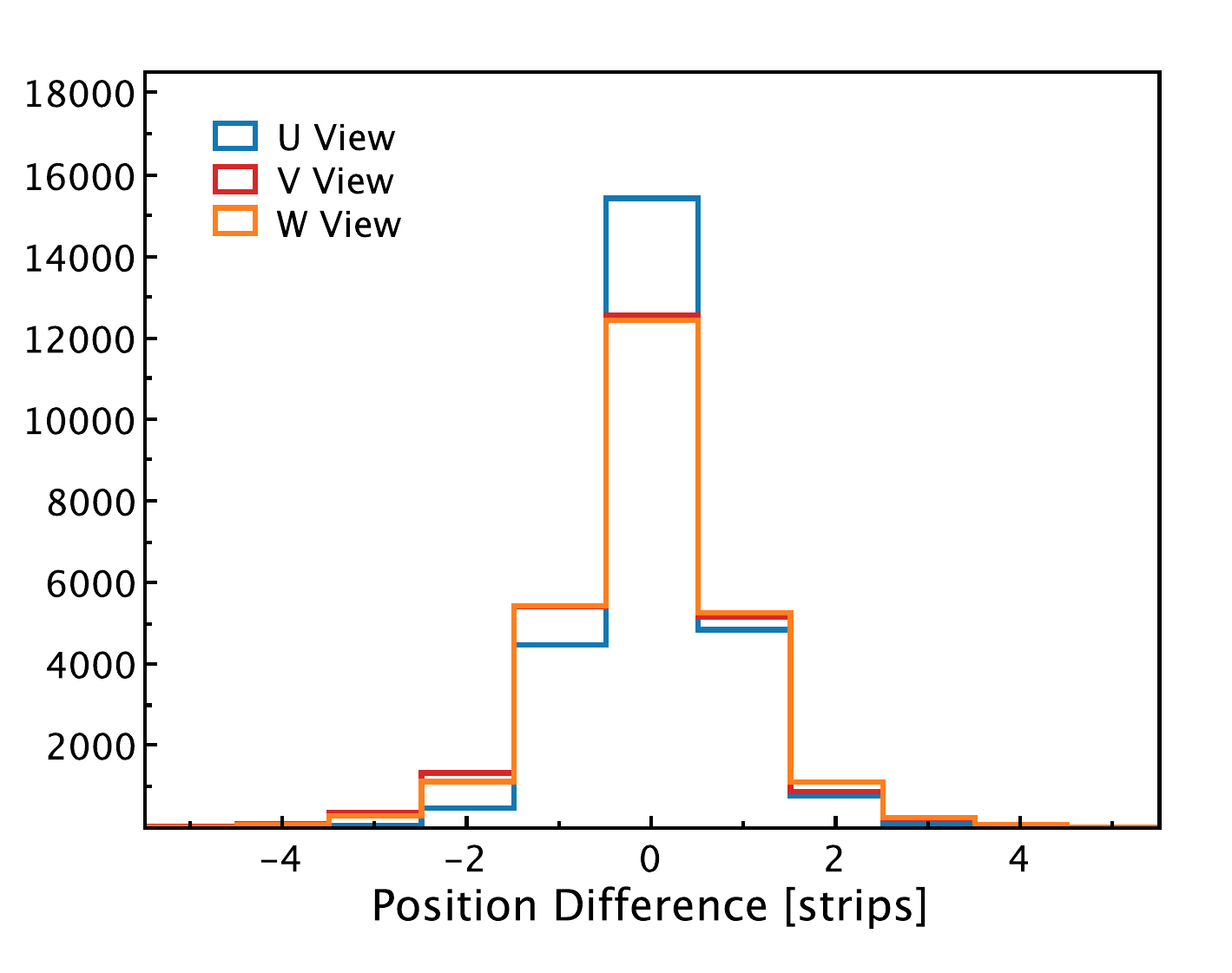}
    \includegraphics[width=0.44\textwidth]{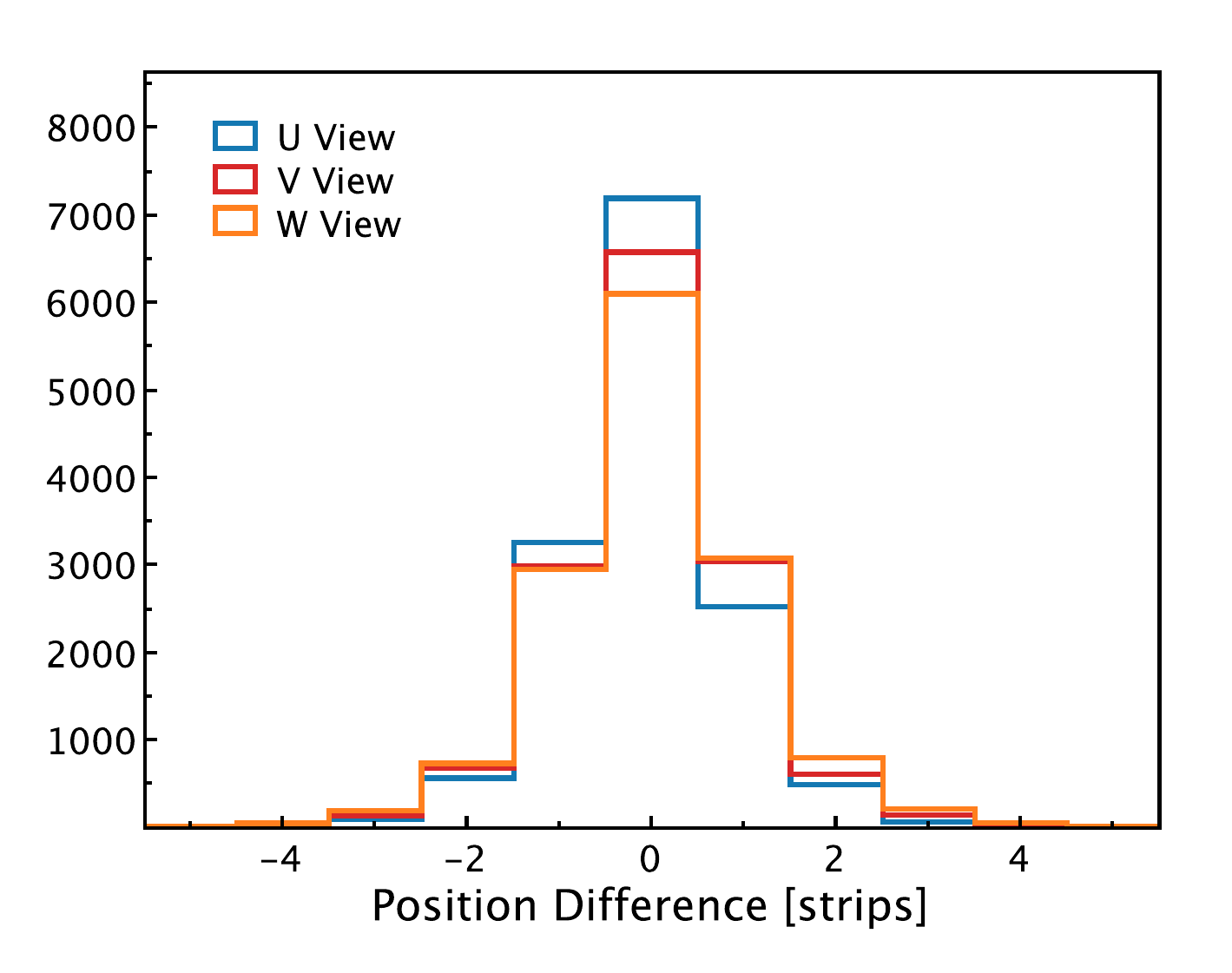}
    \includegraphics[width=0.44\textwidth]{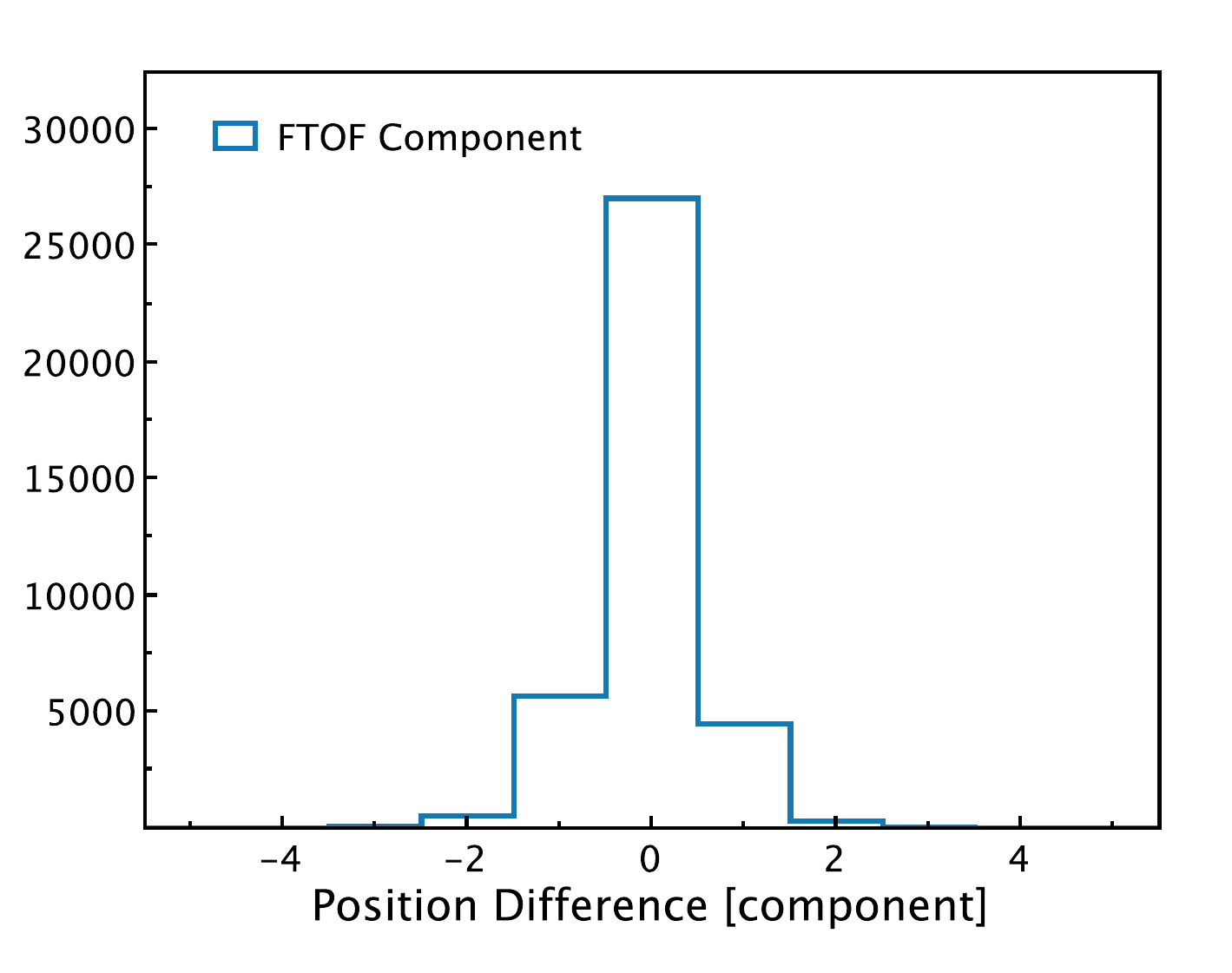}
    \includegraphics[width=0.44\textwidth]{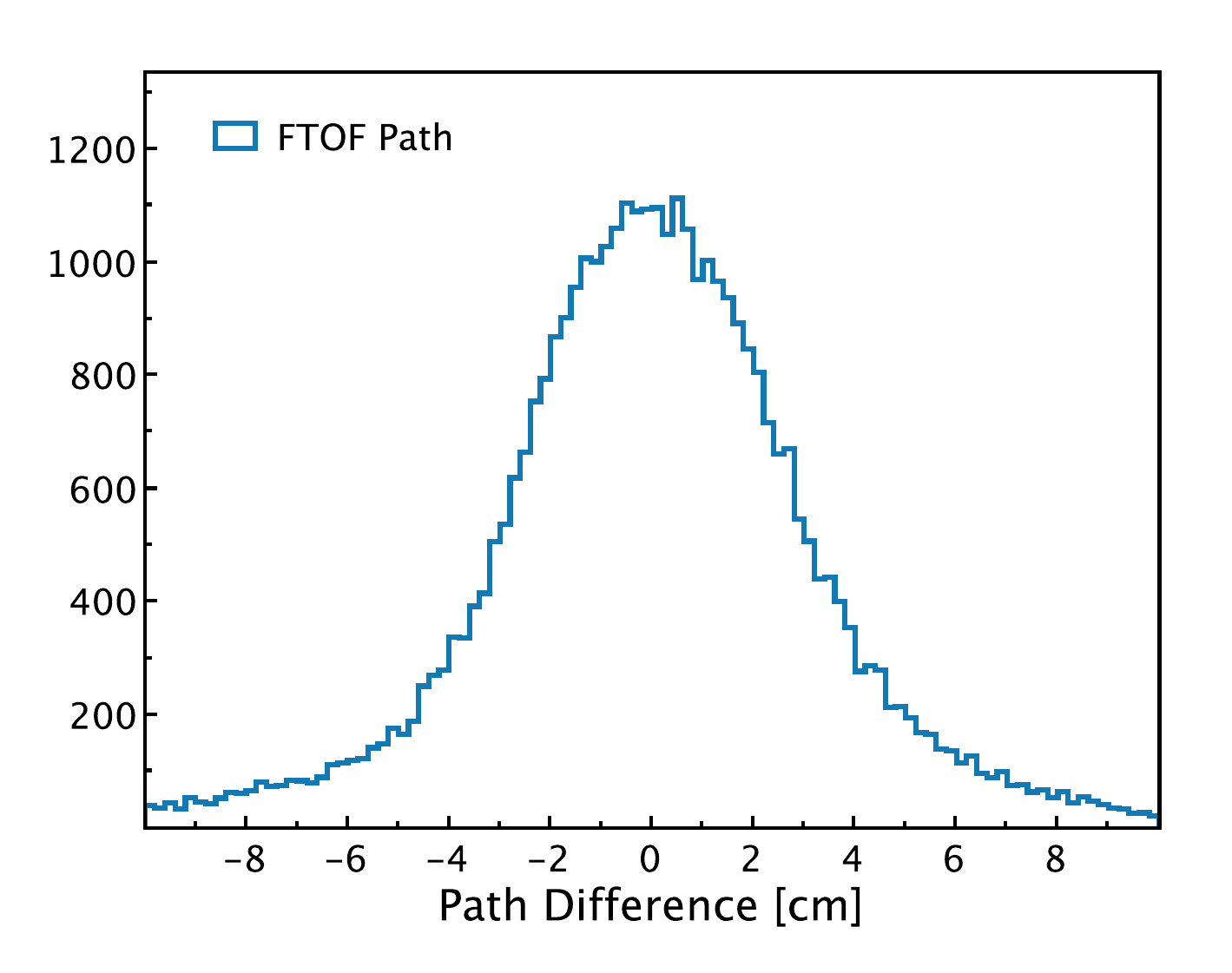}
    \caption[Position Prediction Difference]{The difference between the true and predicted strip positions in the PCAL (top left), ECIN (top middle), ECOUT (top right), FTOF (bottom left) and difference between the true and predicted FTOF path (bottom right). The ECAL position differences are also shown for each of U/V/W views.}
    \label{fig:pred_pos}
\end{figure}

\begin{figure}[ht!]
    \centering   
    \includegraphics[width=0.99\textwidth]{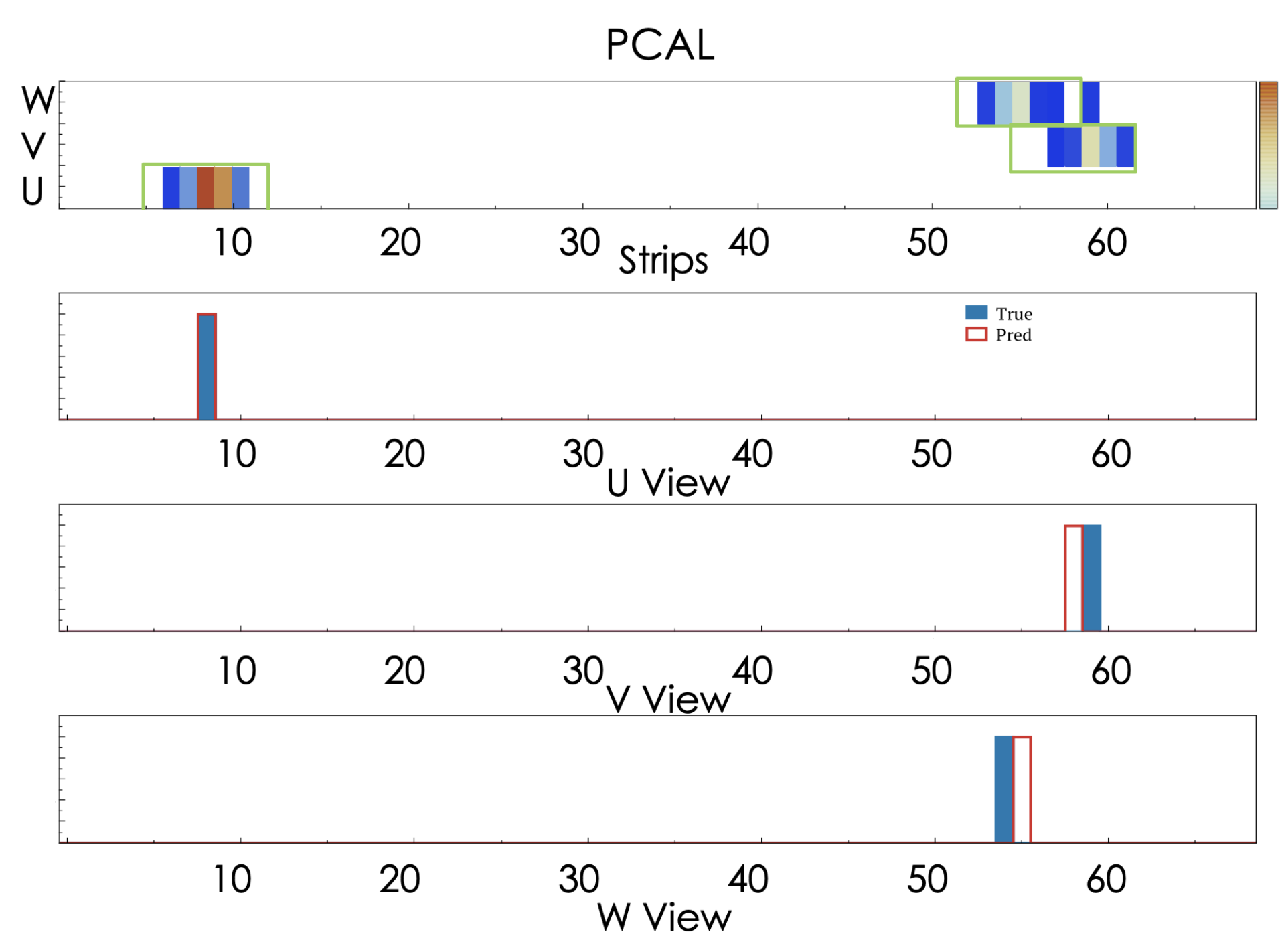}
    \caption[True PCAL Clusters compared to predicted PCAL clusters.]{Top Row: Hits in the U/V/W views of the PCAL. The color axis represents the ADC value for each strip. Green bounding boxes show the strips that would be clustered together based on the predicted cluster positions. Bottom Rows: The true and predicted U, V, W cluster positions.}
    \label{fig:PCAL_Cluster}
\end{figure}

\noindent
Figure~\ref{fig:pred_pos} shows the difference in strips between the predicted position in the PCAL, ECIN, ECOUT and FTOF as well as the difference between the true and predicted FTOF paths. Overall the position difference is mostly within one strip or less, meaning that clusters in the ECAL and FTOF can be easily associated to the correct track by comparing the position of a cluster and the predicted position from the track. An example of how this can be done in practice is shown in Figure~\ref{fig:PCAL_Cluster}. The top row shows the hits in the PCAL whereas the bottom rows show the true and predicted U, V and W cluster positions. A simple clustering algorithm can be made from the predicted cluster positions where any hit within $\pm$ 3 strips is added to the cluster. Note that one hit in W is split off from the cluster due to a strip containing zero energy between the hit and cluster. This hit would not be added to the cluster by the offline reconstruction algorithm and would be interpreted as a neutral particle. As shown in Figure~\ref{fig:PCAL_Cluster} the simple clustering algorithm based on the predicted cluster positions allows to select all hits in the true clusters, ensuring that the correct energy deposition is associated to the track.\\ 

\noindent
Information from the ECAL cluster can then be passed to a neural network to identify the particle that produced the track and cluster as being an electron or not. The Analog-to-Digital Converter (ADC) value associated to each ECAL hit, representing the amplitude of a continuous signal in an ECAL strip, is used in the offline reconstruction to define the hit energy. The online reconstruction will not benefit from the calibration process that allows to convert between ADC value to energy and so will use the raw ECAL ADC values. For the HTCC, the ADC value is proportional to the number of photoelectrons produced by an electron passing through the HTCC. All ECAL variables used for the online identification are: the sum of the ADC value in each of the strips, the position of the cluster and the number of strips hit in each view, for each of the PCAL, ECIN and ECOUT. Information from the HTCC is also passed to the electron identification network. As the HTCC only contains 8 PMTs in a sector, the ADC value from all 8 PMTs are passed as input to the neural network. Clusters in the HTCC sometimes leak into neighboring sectors. To account for this, the ADC value from all 8 PMTs in the neighboring sectors are also given to the electron identification network, leading to the HTCC contributing 24 features. The track's average wire position in each superlayer is also passed to the neural network so that it can correlate the position of HTCC hits to the track. Note that only detector level information such as ADC values are used here to avoid the need for any intermediate reconstruction steps. This amounts to a total of 51 input features. The training sample for the electron identification algorithm is composed from particles identified as electrons by the offline reconstruction in the positive sample, and all other negative particles in the negative sample, for a total of 200000 training events and 60000 testing events). The network has 5 hidden layers with ReLu activation functions and 50, 100, 50, 25 and 5 nodes respectively. The output layer uses a sigmoid activation function. The network is implemented using CLAS12 specific Java based tools used in other machine learning applications at CLAS12~\cite{AI_Tracking,AI_Tracking2}, available via the Maven build automation tool~\cite{maven,mavenRepo}.\\

\begin{figure}[ht!]
    \centering   
    \includegraphics[width=0.49\textwidth]{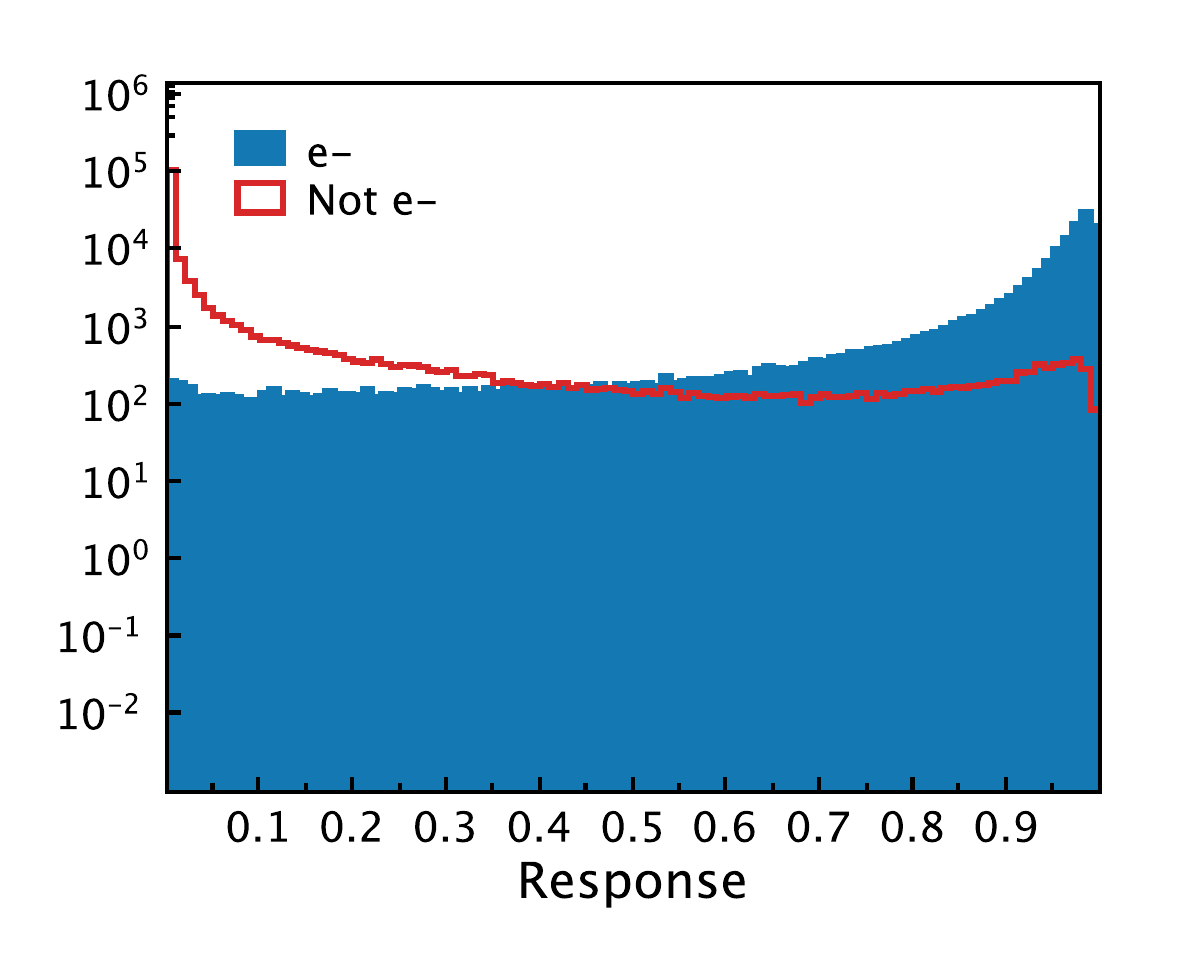}
    \includegraphics[width=0.49\textwidth]{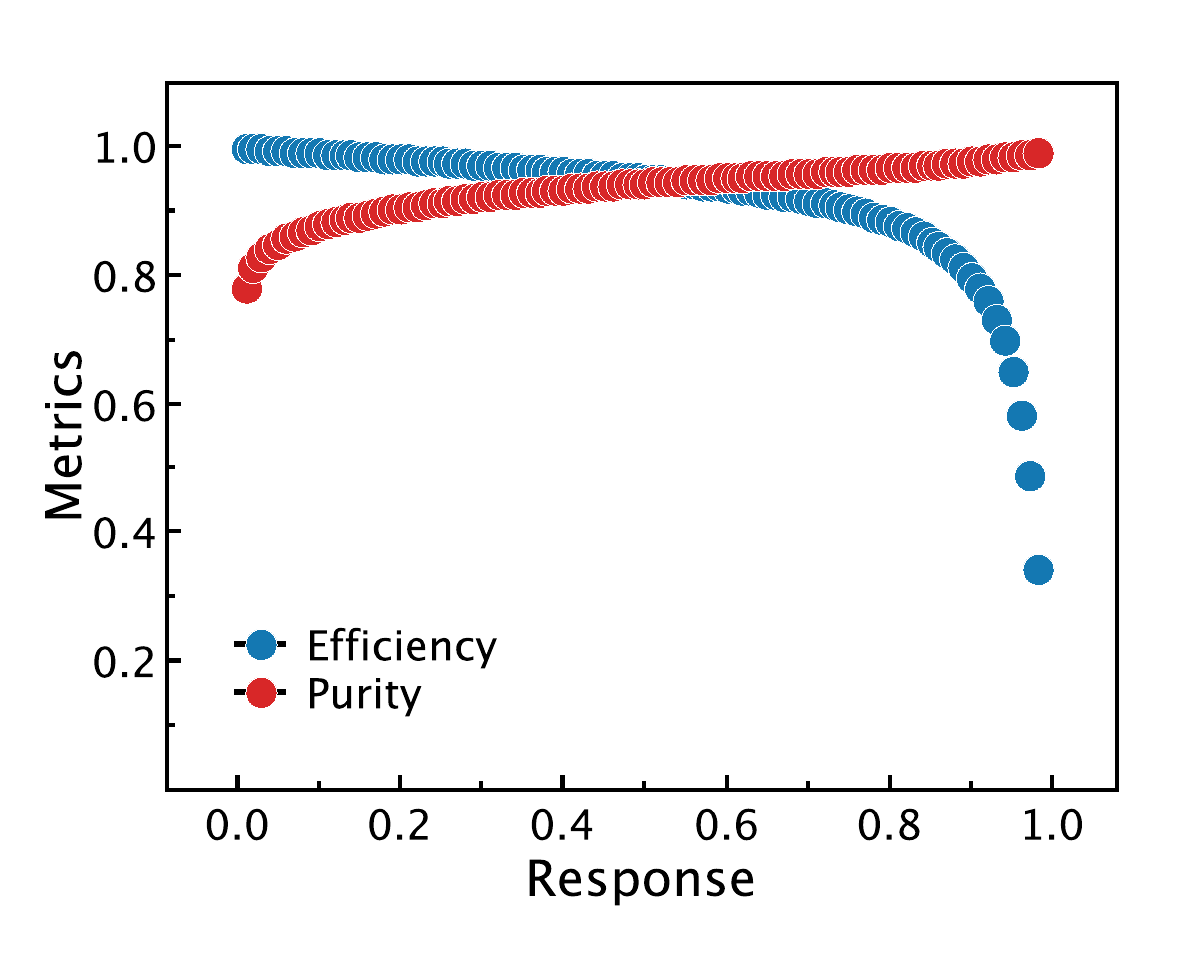}
    \caption[The training response and metrics]{The training response (left) and the efficiency and purity (right) of the electron identification network.}
    \label{fig:train_mets}
\end{figure}

\noindent
Figure~\ref{fig:train_mets} shows the output of the network for both positive and negative samples where a value of 1 in the training data is assigned to electrons and 0 to all other particles. This output is called the response. The response is used to decide if the network labels a particle as an electron or not. A threshold is placed on the response whereby any particle with a response above the threshold is labeled an electron and any particle with a response below the threshold is labeled as not being an electron. As expected, the response peaks at 1 for electrons and at 0 for other particles. There is however a smaller increase at 1 for other particles, which is indicative that some particles are either incorrectly labeled as not being electrons in the training data or are incorrectly labeled as being electrons by the network. This will be discussed in more detail in the next section. Figure~\ref{fig:train_mets} also shows the performance of the network in terms of electron identification efficiency (E) and purity (P) calculated as:

\begin{equation} E=\frac{TP}{TP+FN} \end{equation}
\\
\begin{equation} P=\frac{TP}{TP+FP} \end{equation}

\noindent
where TP is the number of electrons correctly identified by the network (True Positives), FP is the number of non electrons incorrectly labeled as electrons by the network (False Positives) and FN is the number of electrons incorrectly labeled as non electrons by the network (False Negatives). The statistical uncertainties on the efficiency and purity, although negligible here due to large datasets, are calculated as described in Ref.~\cite{Unc}. The efficiency and purity are plotted as a function of the threshold placed on the response. This threshold can then be varied based on the objective at hand. If used for triggering, a high efficiency is desirable to avoid discarding good data. In this case the threshold would be placed at a low value of the response. For precise analysis, low backgrounds in electron identification may be desirable, and the threshold on the response would then be placed at a higher value. Overall Figure~\ref{fig:train_mets} demonstrates a good performance of the electron identification algorithm, with an efficiency above 99.5\% preserved by a cut at 0.1 or below on the response. Note, however, that the purity here is not the purity that would be attainable online, due to the fact that here this is calculated for equal amounts of electrons and other particles, whereas in reality this is not the case.\\

\begin{figure}[ht!]
    \centering   
    \includegraphics[width=0.85\textwidth]{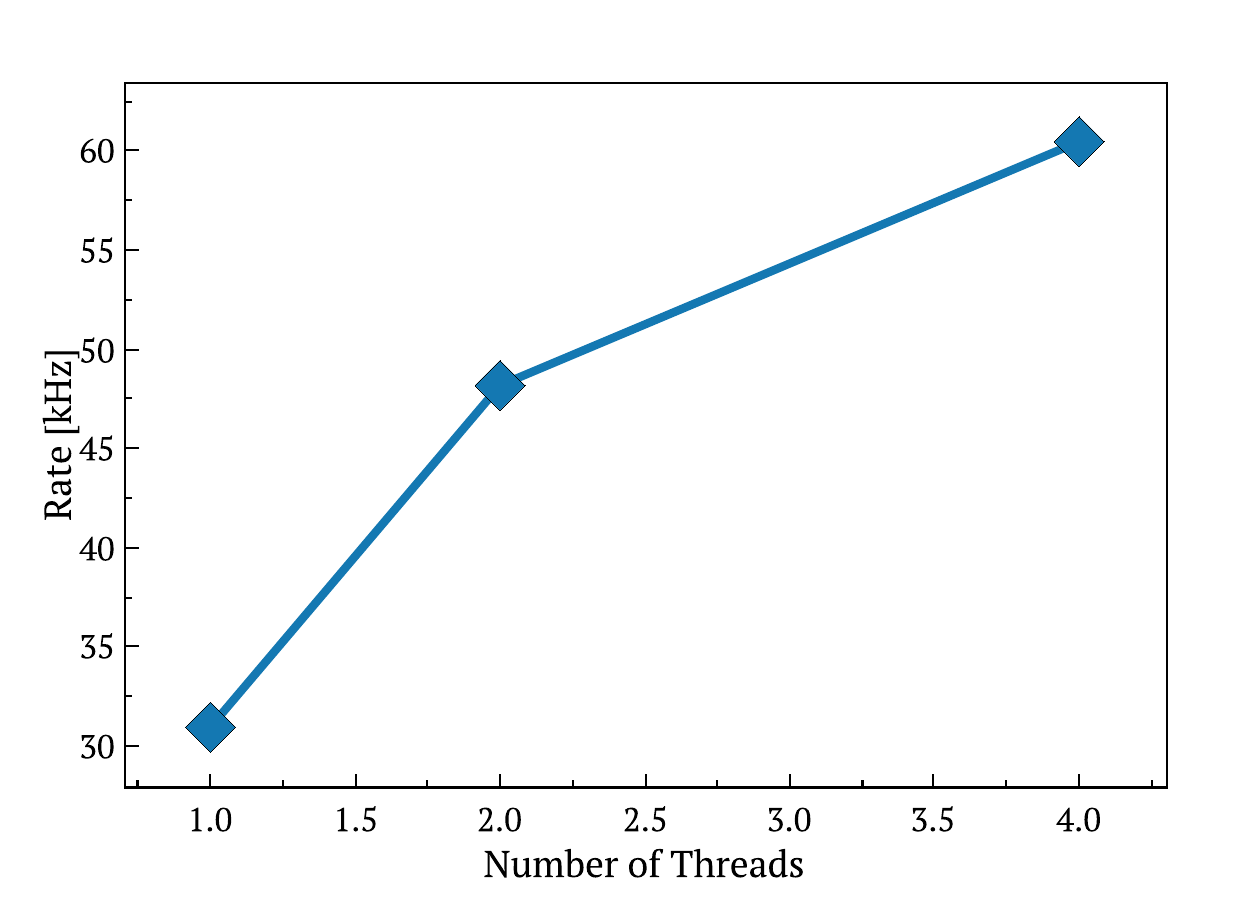}
    \caption[The electron identification prediction rate]{The electron identification prediction rate as a function of threads (CPU cores) for a 2.4 GHz 8-Core Intel Core i9 MacBook Pro.}
    \label{fig:rates}
\end{figure}

\noindent
A final important consideration is the prediction rate for this electron identification scheme. Typical trigger rates at CLAS12 are on the order of 20 kHz~\cite{C12Software,C12Trigger} and there are plans to increase the data taking luminosity which would increase the trigger rates. The online electron identification scheme proposed here would run after the conventional electron trigger. As such this identification scheme needs to at least match the electron trigger rate. Figure~\ref{fig:rates} shows the combined prediction rate of the cluster finding and electron identification networks as a function of the number of CPU cores as measured on a 2.4 GHz 8-Core Intel Core i9 MacBook Pro. The prediction rate roughly scales as a function of CPU cores. Even with a single core the prediction rate for these two networks is high enough to match the electron trigger rate. Note that this is also much higher than the convolutional neural network based electron trigger scheme proposed in Ref.~\cite{AITrigger}. The overall online processing rate will be slower than shown in Figure~\ref{fig:rates}, due to for example the tracking described at the start of this section, but will be deployed on many CPU cores which will allow to match the data taking rate.

\section{Validation} \label{sect_tests}

\noindent
As mentioned at the end of Section~\ref{sect_algo}, the purity and efficiency of the online electron identification algorithm should be measured based on all negative particles in CLAS12 data. To obtain an estimation of the purity as fair as possible, the only restriction on the offline electron identification is that the vertex of the electron is restricted to a reasonable range around the location of the target. To avoid any confusion between sources of inefficiency, the comparison between online and offline identification is made for tracks that are reconstructed both online and offline. That is to say that the efficiency and purity measured here only relate to the online and offline electron identification algorithms and not track reconstruction algorithms. The tests were made on data taken with a liquid hydrogen target at a beam current of 45 nA which was the typical running conditions at CLAS12 when the dataset was taken.\\

\begin{figure}[ht!]
    \centering   
    \includegraphics[width=0.45\textwidth]{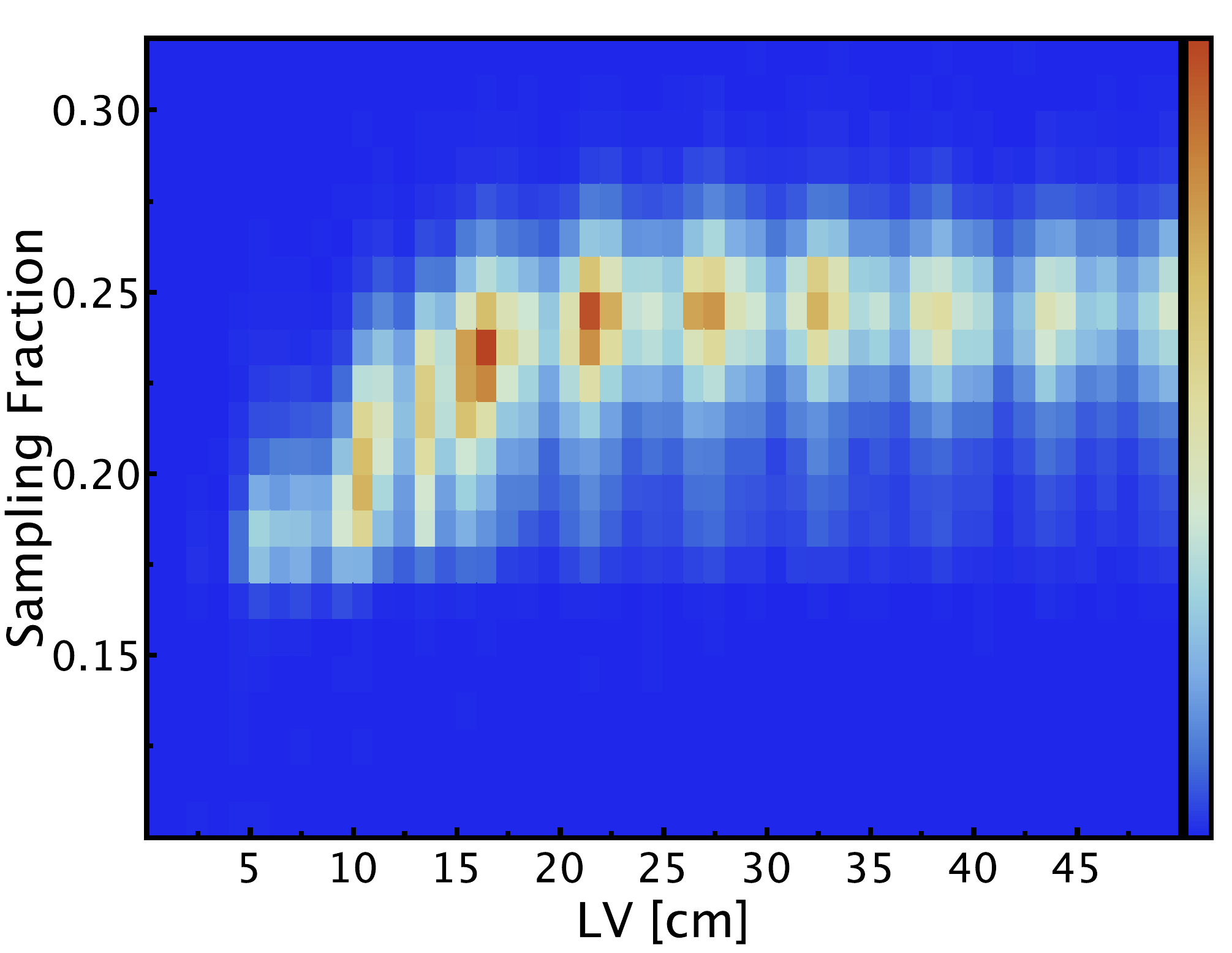}
    \includegraphics[width=0.45\textwidth]{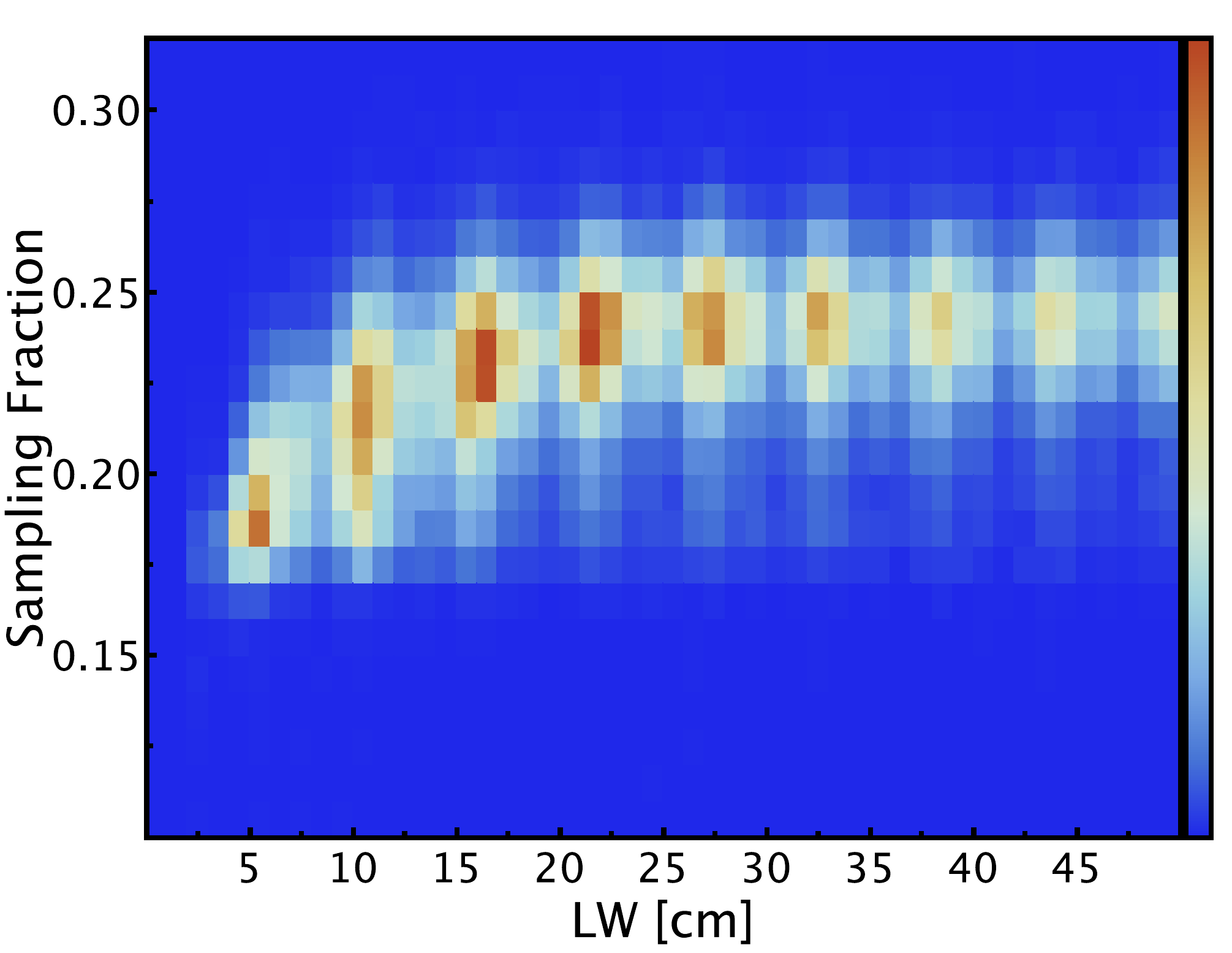}
    \caption[The electron sampling fraction as a function of local position.]{The electron sampling fraction as a function of local position in the V and W views. Loose and tight fiducial cuts are placed at at 9 and 13.5 cm.}
    \label{fig:SF_L}
\end{figure}

\noindent
The easiest measure of the purity and efficiency of the online reconstruction is relative to the offline reconstruction. However, the offline electron identification algorithm is known to have some sources of inefficiencies. For example, there are some issues in the HTCC clustering and associating HTCC cluster to tracks suffers from resolution effects at low angles. Another issue is due to the cuts on the electron sampling fraction. Figure~\ref{fig:SF_L} shows the electron sampling fraction as a function of local position in the V and W views of the PCAL. As seen the sampling fraction drops as the electron gets closer to the edge of the calorimeter. This is due to the fact that the electromagnetic shower produced by the electrons may not be fully contained within the PCAL when electrons hit close to its edge. Fiducial cuts are typically placed at a distance of 2 strips (each 4.5cm wide) from the edge of the calorimeter to ensure that the shower is fully contained within the PCAL. Tighter cuts can be placed at a distance of 3 strips. The electron sampling fraction is decreased outside of the fiducial region of the calorimeter selected by the cuts. This causes issues with the offline electron identification which misses those electrons with a low sampling fraction.\\

\begin{figure}[ht!]
    \centering   
    \includegraphics[width=0.75\textwidth]{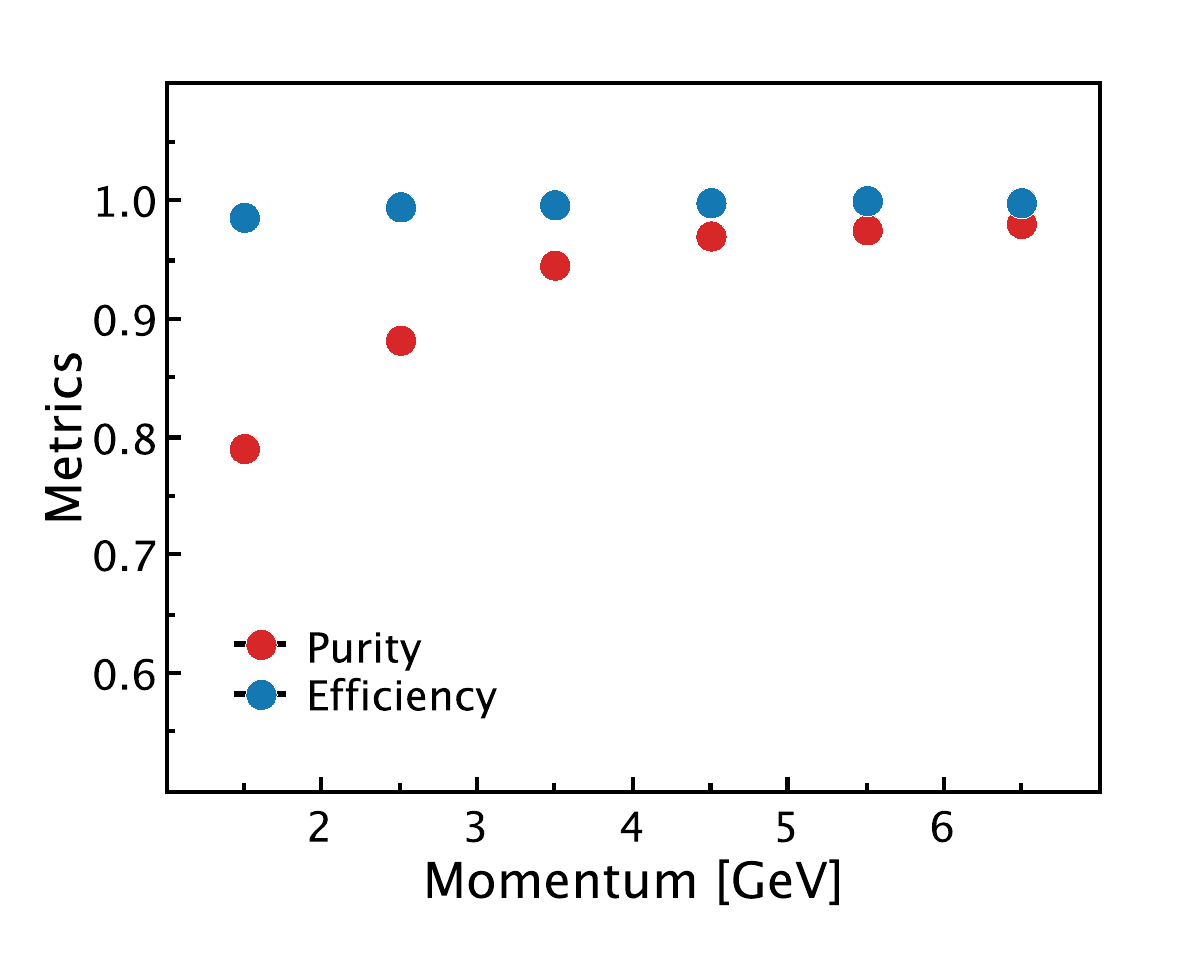}
    \caption[The online electron identification efficiency and purity.]{The online electron identification efficiency and purity relative to the offline electron identification algorithm.}
    \label{fig:mets_p}
\end{figure}

\noindent
Figure~\ref{fig:mets_p} shows the online electron identification efficiency and purity relative to the offline electron identification algorithm, within 2 strips of the edge of the calorimeter as shown in Figure~\ref{fig:SF_L}. The threshold on the response of the online electron identification algorithm was here placed at 0.075. As shown the electron identification has an efficiency close to 100\%. The purity is also above 75\% and reaching 90\% and above at high momentum. The online electron identification found at least one electron in only 43\% of events where the conventional electron trigger with DC roads found at least one electron. Assuming an electron identification efficiency of 100\% as shown in Figure~\ref{fig:mets_p}, the online electron identification could be used as a second pass filter after the conventional electron trigger and achieve a data reduction of 57\% in these running conditions. This is comparable to what could be achieved with the CNN based approach of Ref.~\cite{AITrigger} described in the introduction. It must however be stressed that the method presented here has the added benefit of providing reconstructed electrons for further analysis.\\

\begin{figure}[ht!]
    \centering   
    \includegraphics[width=0.8\textwidth]{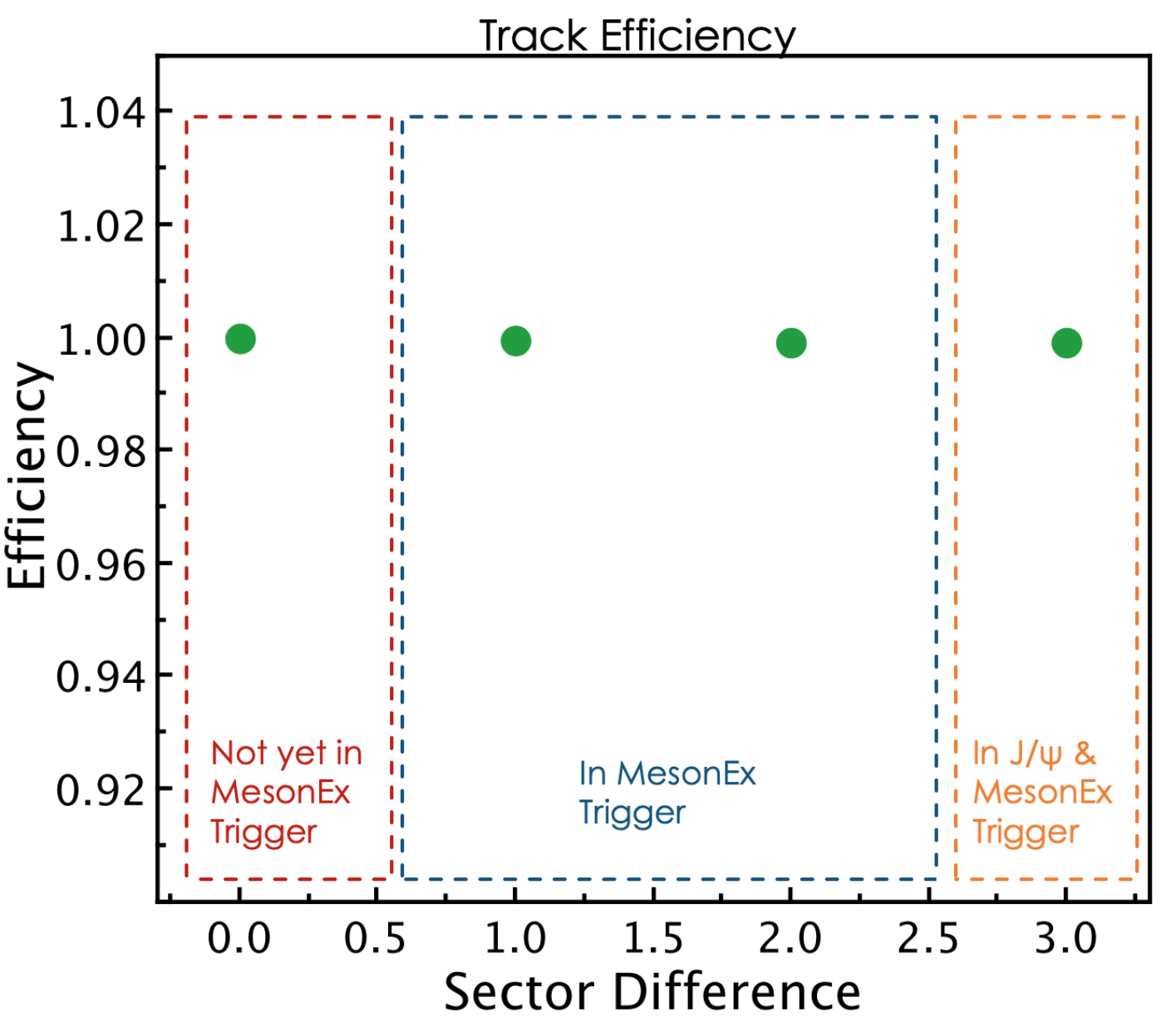}
    \caption[The efficiency of the online reconstruction schemes at finding two tracks with a given separation in sectors.]{The efficiency of the online reconstruction schemes at finding two tracks with a given separation in sectors. The bounding boxes show which sector separation are currently or yet to be implemented in the conventional MesonEx or J/$\psi$ triggers.}
    \label{fig:nonElTrigger}
\end{figure}

\noindent
Consideration should also be extended to the photoproduction (so called MesonEx) and J/$\psi$ triggers~\cite{C12Trigger} described at the end of Section~\ref{sect_conv}. The MesonEx trigger requires two charged hadron tracks in the FD. The conventional trigger cannot accommodate this trigger when two tracks are in the same sector. The J/$\psi$ trigger requires two minimum ionising particles (MIPs) in opposite sectors. MIP-candidates are identified as having low energy deposition in the calorimeters. Although no online MIP identification algorithm has been developed, the current scheme can at least reconstruct two tracks in opposite sectors that do not belong to electrons. These tracks must either belong to muons or hadrons, with hadrons being predominantly produced over muons. Figure~\ref{fig:nonElTrigger} shows the efficiency in recovering two hadron or muon tracks as a function of the number of sectors separating the two tracks. The efficiency here measures the number of events with at least two tracks with a given sector difference in the online reconstruction for events with at least two tracks with the same sector difference in the offline reconstruction. As shown, the current scheme would allow to accommodate the MesonEx and J/$\psi$ triggers whilst retaining the same efficiency relative to the offline reconstruction algorithm.\\

\noindent
Exclusive reactions can be used to identify electrons without having to rely on the offline identification scheme. For example, one can look at the reaction $ep \to e'\pi^+\pi^-(p)$  where the proton is undetected and the scattered electron is written as e'. The events produced in this reaction can be selected using the missing mass of $ep \to e' \pi^+\pi^- X$ that will peak at the mass of the proton (0.94 GeV). The number of $ep \to e'\pi^+\pi^-(p)$  events is counted by fitting the missing mass distribution. The positively charged pion is identified using the offline particle identification scheme. The negatively charged pion candidate is not identified using the offline reconstruction and just required to be associated with a negatively charged track. The electron candidate tracks are required to have at least a hit in one of the calorimeters, and the HTCC is also required to have a hit in the same or neighboring sectors, although the hit does not have to be associated to the electron. The missing mass of $ep \to e' \pi^+\pi^- X$ will only peak at the proton mass if the correct mass hypothesis is assigned to each final state particle. The presence of the proton mass peak ensures that the final state particles were assigned the correct mass, and therefore ensures that the electron and pion candidates are truly an electron and pion respectively. The electron identification efficiency can then be obtained by comparing the total number of $ep \to e'\pi^+\pi^-(p)$  events to those where the electron candidate is identified as an electron by either the offline or online reconstruction schemes. Note that the tracks are required to be matched in the offline and online reconstructions, ensuring that the tracks are found in both the offline and online reconstruction such that the efficiency only takes into account the particle identification efficiency. \\

\begin{figure}[ht!]
    \centering   
    \includegraphics[width=0.49\textwidth]{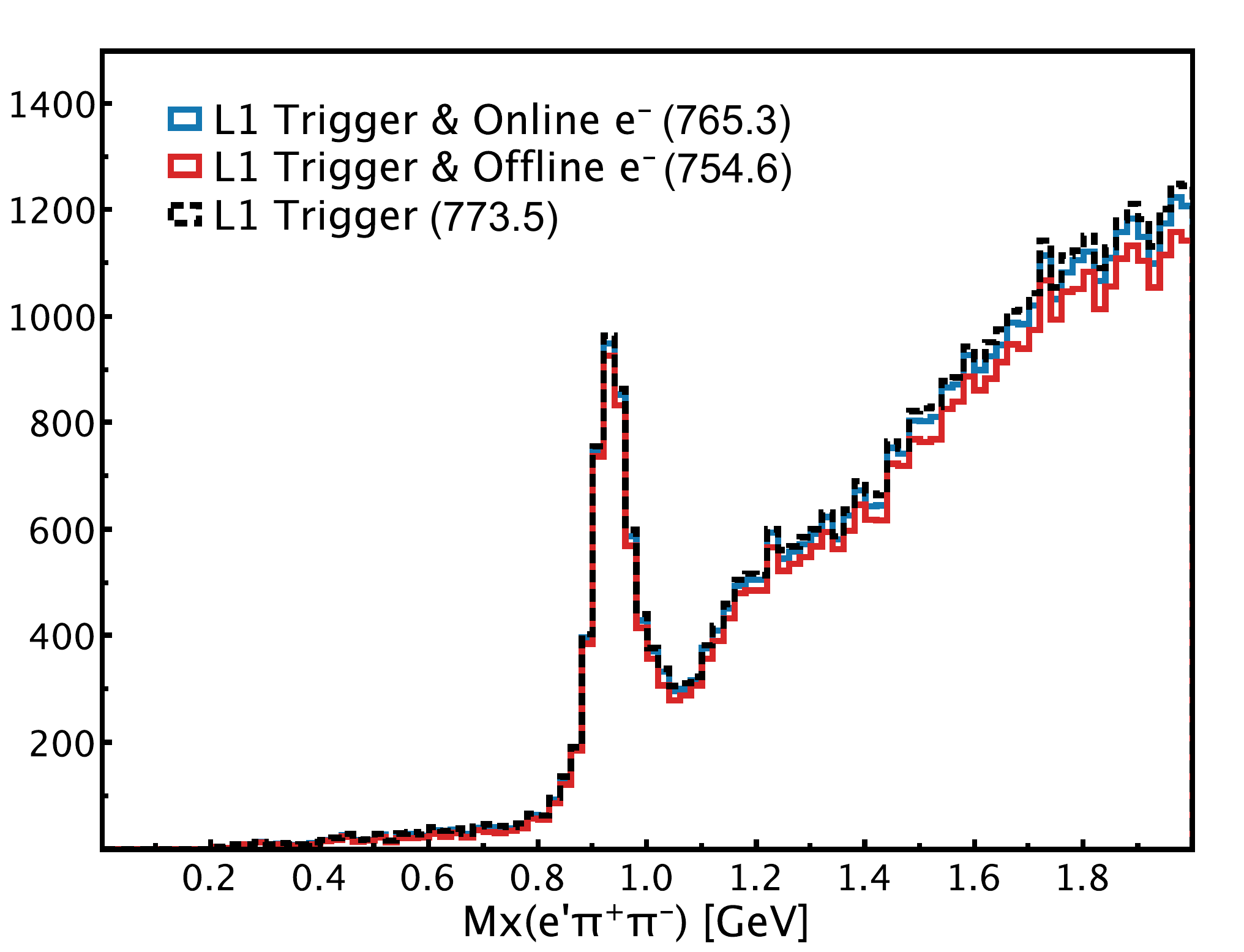}
    \includegraphics[width=0.49\textwidth]{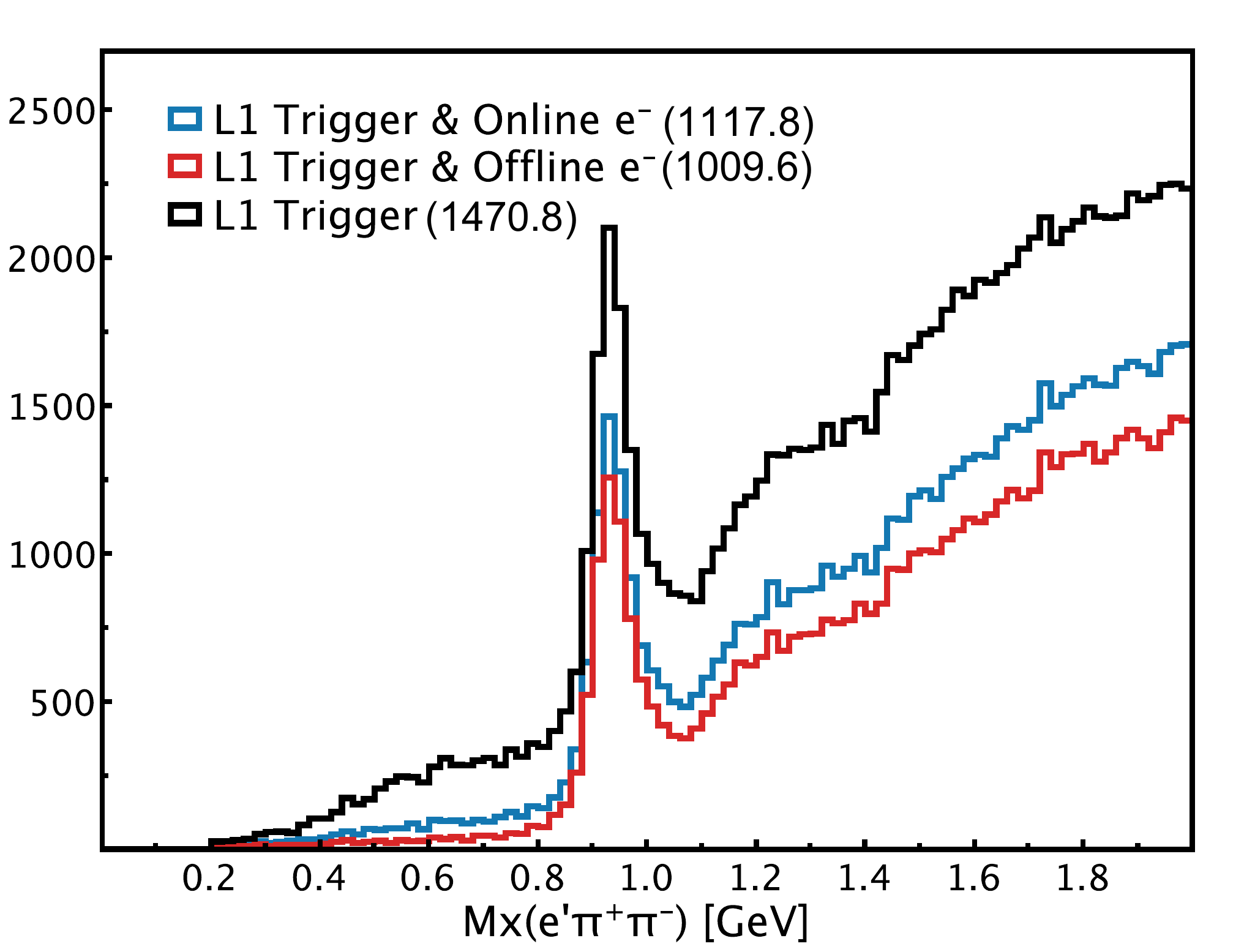}
    \caption[The missing mass of $ep \to e' \pi^+\pi^- X$.]{The missing mass of $ep \to e' \pi^+\pi^- X$ for all events selected by the conventional level 1 electron trigger (black) and events where the electron candidate was identified as an electron by the offline (red) and online (blue) identification schemes. The missing mass distributions are plotted with cuts on the distance from the edge of the PCAL at 13.5 cm (left) and without the cuts (right). The number of electrons estimated from the fit to the proton peak in the missing mass distributions are shown within the legend.}
    \label{fig:MM_n}
\end{figure}

\begin{figure}[ht!]
    \centering   
    \includegraphics[width=0.9\textwidth]{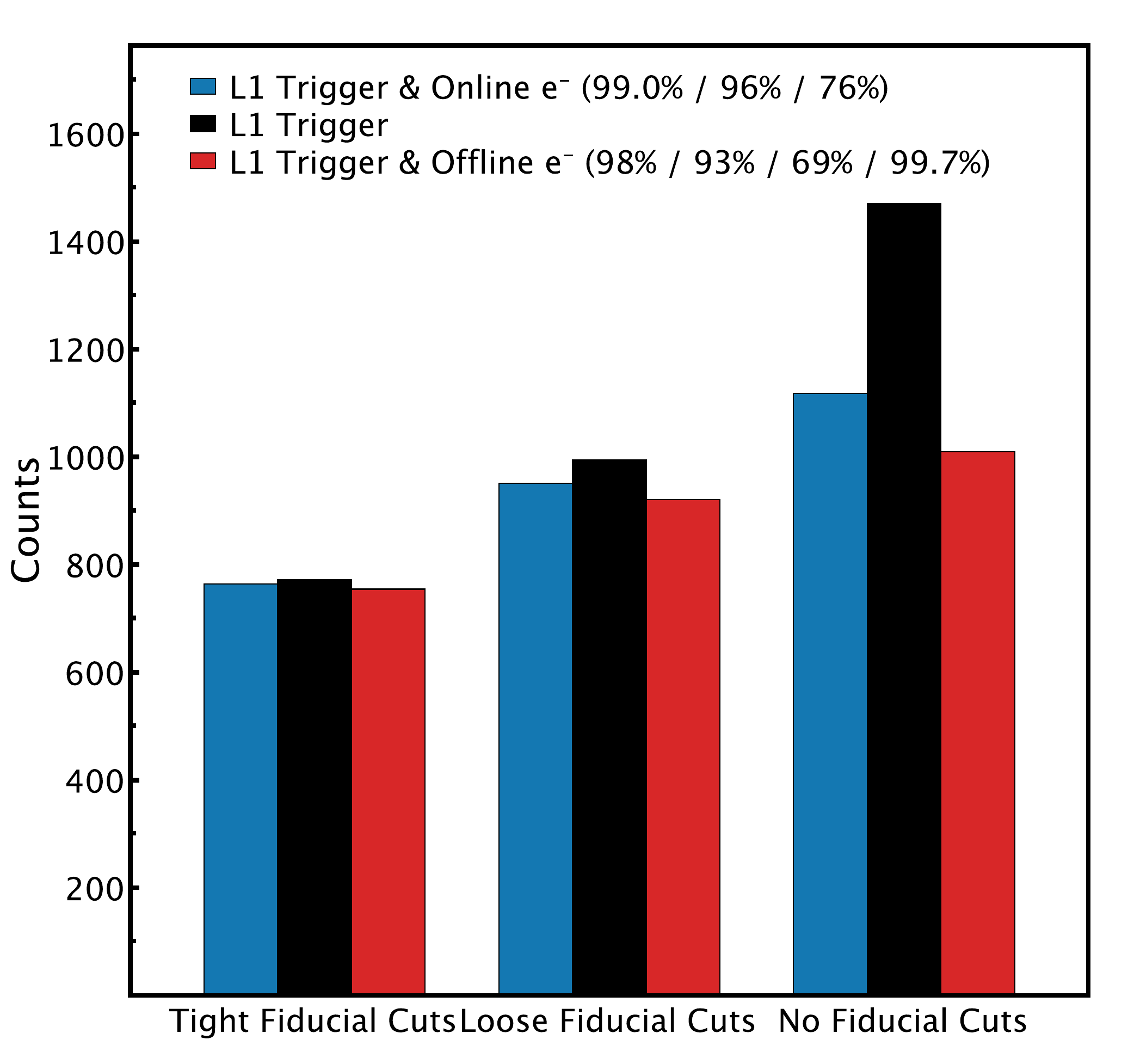}
    \caption[The number of electrons in $ep \to e' \pi^+\pi^- X$ events and identified by the online and offline schemes.]{The total number of $ep \to e' \pi^+\pi^- X$ events selected by the conventional level 1 trigger (black) and those where the electron is correctly identified by the online (blue) and offline (red) schemes, for the tight fiducial cuts at 13.5 cm, the looser fiducial cuts at 9 cm and without fiducial cuts. }
    \label{fig:MM_n_barchart}
\end{figure}

\noindent
Figure~\ref{fig:MM_n} shows the missing mass of $ep \to e' \pi^+\pi^- X$ for all events selected by the conventional level 1 electron trigger and events where the electron candidate was identified as an electron by the offline or online identification schemes. Figure~\ref{fig:MM_n} also highlights the impact of the cuts on the fiducial region of the PCAL, with cuts on the distance from the edge of the PCAL at 13.5 cm and without the cuts. The number of electrons estimated by fitting the missing mass distributions are shown in Figure~\ref{fig:MM_n_barchart}. As shown, the online and offline electron identification schemes reach efficiencies close to 100\% when applying fiducial cuts. However, as the cuts are loosened, first to 9cm then removed, the efficiency of the identification schemes decrease. This is due to the drop in the sampling fraction close to the edge of the calorimeter, which then leads to electrons being rejected by the offline particle identification scheme. As the offline scheme is used to create the training sample for the online scheme, it will also not correctly identify electrons close to the edge of the calorimeter as these were not present or incorrectly labeled as non-electrons in the training sample. The small increase in efficiency of the online electron scheme is attributed to the fact that it does not have a cut on the sampling fraction and therefore correctly identifies some electrons that would just be rejected by the offline identification scheme. This leads to the peak in the response close to 1 for non-electrons as shown in Figure~\ref{fig:train_mets} and highlighted in the previous section.\\

\noindent
There are two conclusions to be drawn from Figure~\ref{fig:MM_n_barchart}. First, the offline identification scheme is not suitable to estimate the absolute efficiency of the online scheme, as the offline identification scheme is not fully efficient. The purity estimated relative to the offline identification scheme is also lower than the true purity, as the online identification scheme is able to preserve electrons that are lost by the offline reconstruction. Secondly, at present the online electron identification would not be fully suitable as a second pass trigger as it is not perfectly efficient and would therefore decrease the statistics available for analysis. This is especially true for electrons close to the edge of the calorimeters, although the online identification scheme is nearly fully efficient for electrons within the fiducial region of the calorimeters. The next section will describe how training samples can be created without fully relying on the offline identification which will improve the electron identification.\\

\begin{figure}[ht!]
    \centering   
    \includegraphics[width=0.49\textwidth]{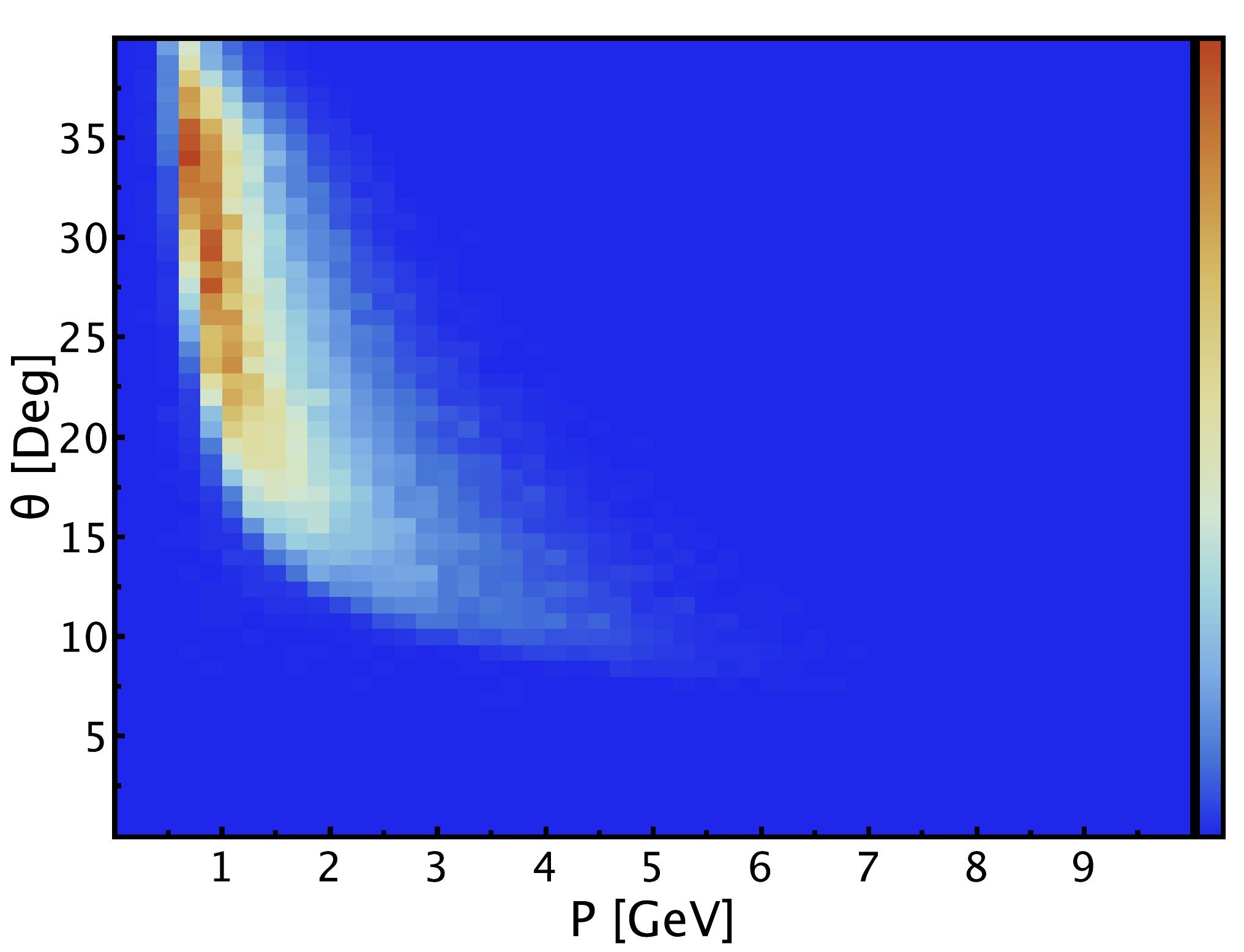}
    \includegraphics[width=0.49\textwidth]{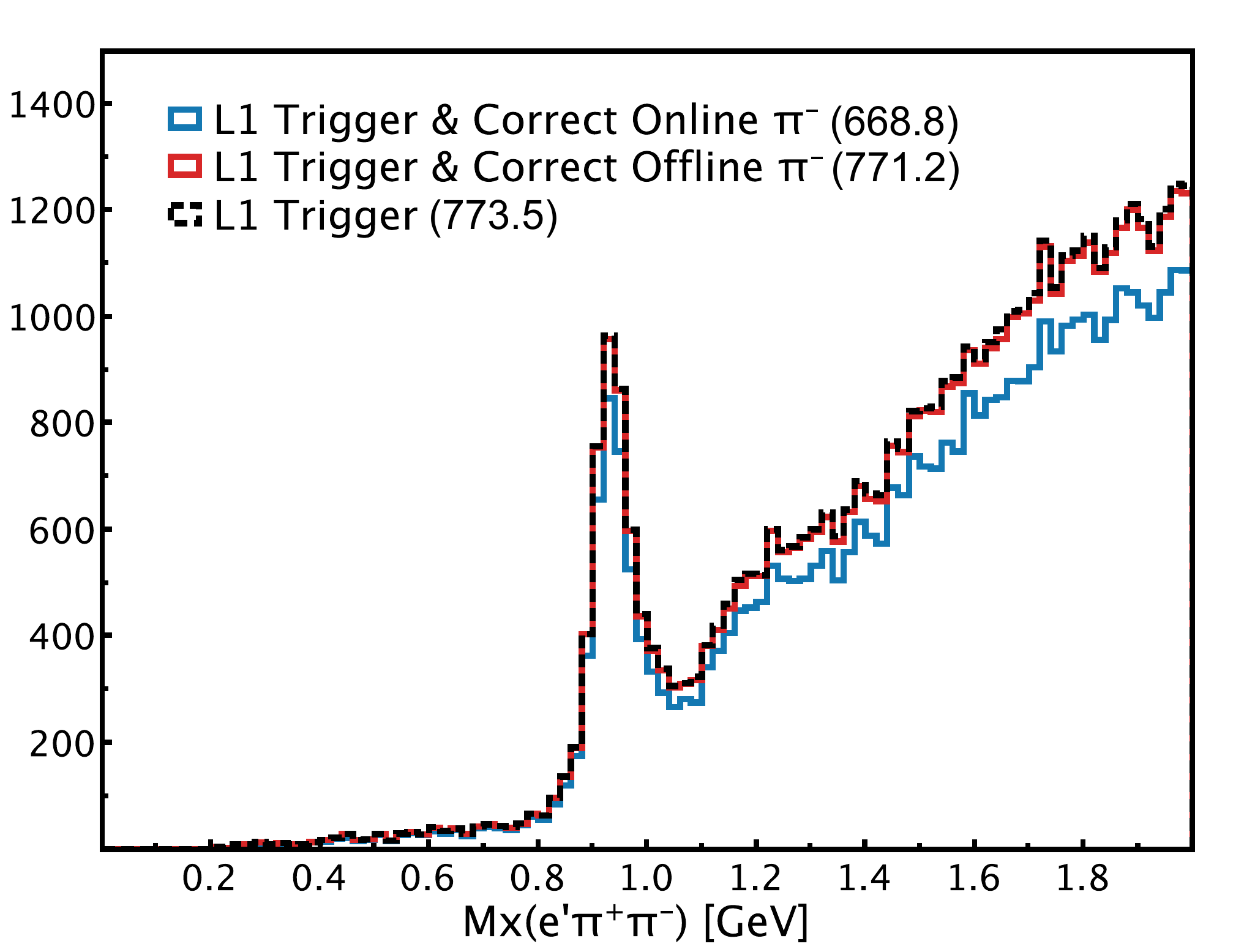}
    \caption[The momentum of pions in events selected with the missing mass of $ep \to e' \pi^+\pi^- X$.]{Left: The polar angle vs the momentum of negatively charged pions in the missing range of $ep \to e' \pi^+\pi^- X$ from 0.85 to1.0 GeV. Right: The missing mass of $ep \to e' \pi^+\pi^- X$ for all events selected by the conventional level 1 electron trigger (black) and events where the pion candidate was \textbf{not} identified as an electron by the offline (red) and online (blue) identification schemes. The missing mass distributions are plotted with cuts on the distance from the edge of the PCAL at 13.5 cm. The number of electrons estimated from the fit to the proton peak in the missing mass distributions are shown within the legend.}
    \label{fig:MM_pur}
\end{figure}

\noindent
A similar argument can be made relating to the electron identification purity. As shown in Figure~\ref{fig:HTCC}, pions above 4.5 GeV start to produce hits in the HTCC. These pions are then identified as electrons, decreasing the electron identification purity in the offline reconstruction. As the online identification scheme is trained using the offline identification to create training samples, the same issues surrounding the electron purity will be present in the online identification scheme. The electron purity can be estimated by comparing the total number of $ep \to e'\pi^+\pi^-(p)$ events to those where the pion candidate is \textbf{not} identified as an electron by the offline or online reconstruction schemes. Note that here the concern is not that the pion candidate is identified or not as a pion, which would require distinguishing between different hadron types, but simply that it is not identified as an electron as this would decrease the electron identification purity. Figure~\ref{fig:MM_pur} shows the missing mass of $ep \to e' \pi^+\pi^- X$ for all events selected by the conventional level 1 electron trigger and events where the pion candidate was correctly \textbf{not} identified as an electron by the offline and online identification schemes. As shown, the offline reconstruction has very few events where the pion candidate is incorrectly identified as an electron. This is due to the kinematics of the negatively charged pion which mostly has momentum below 4.5 GeV, also shown in Figure~\ref{fig:MM_pur}. In that momentum range pions mostly do not fire the HTCC, and therefore are not identified by the offline reconstruction as electrons. The online electron identification purity is decreased relative to the offline electron purity by about 87\% which is consistent with the estimate obtained in Figure~\ref{fig:mets_p}.\\

\noindent
Unfortunately, it is hard to find well resolvable samples of pions at high momentum where the pion identification can be validated using exclusive reactions. At momenta above 4.5 GeV, one would expect the offline electron purity to be less than 100\%. The primary objective for the electron identification scheme presented in this article is to achieve comparable performance to the offline identification scheme whilst being fast enough to keep up with the data taking rate, such that it can be employed for online analysis during data taking operations. A high efficiency is further desirable as it would allow for online event selection and triggering. The purity of the online electron identification scheme will not be investigated further, given that it is not a priority here, and will be quoted relative to the offline identification.\\

\section{Improving the Electron Identification Efficiency} \label{sect_retrain}
\begin{figure}[ht!]
    \centering   
    \includegraphics[width=0.45\textwidth]{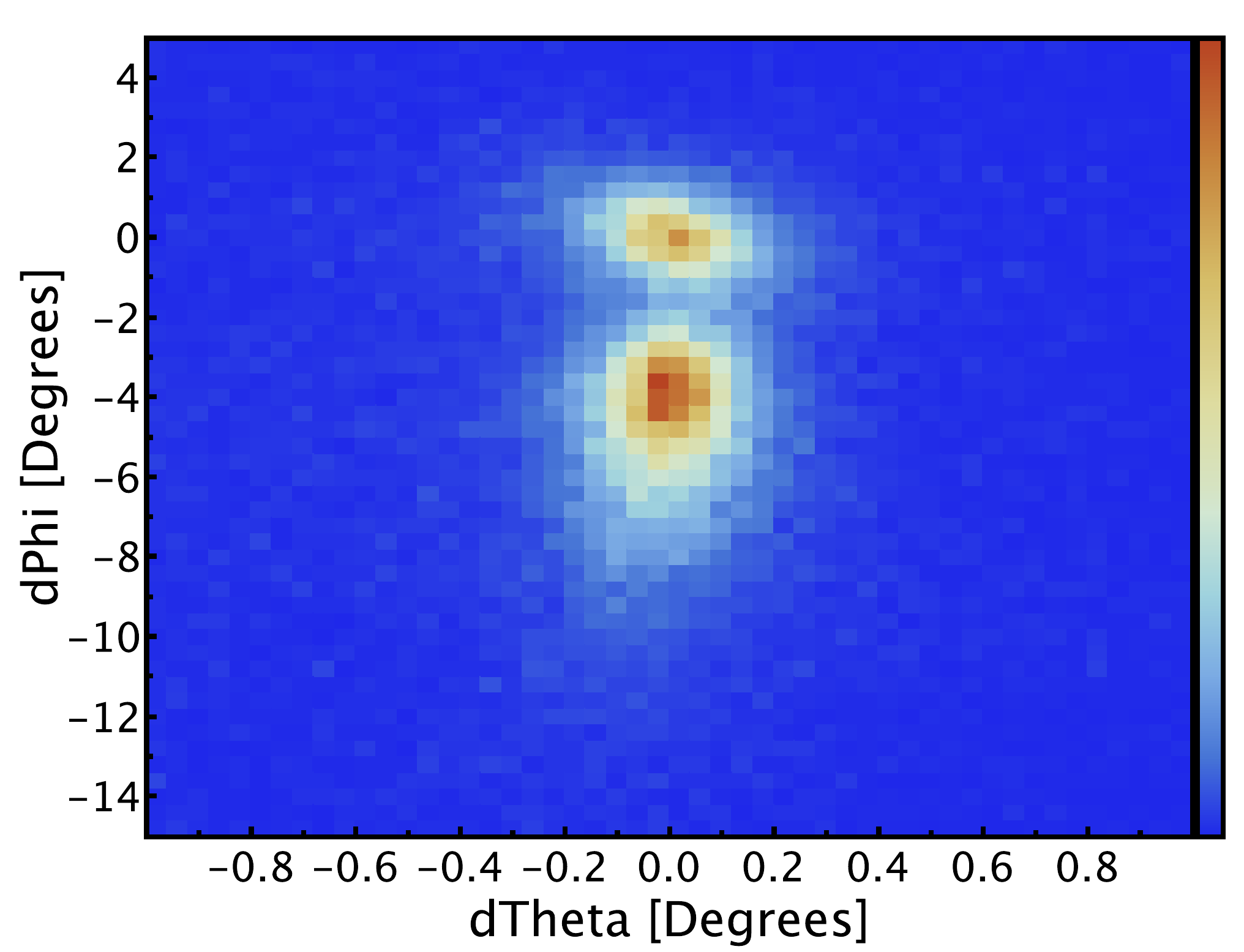}
    \includegraphics[width=0.45\textwidth]{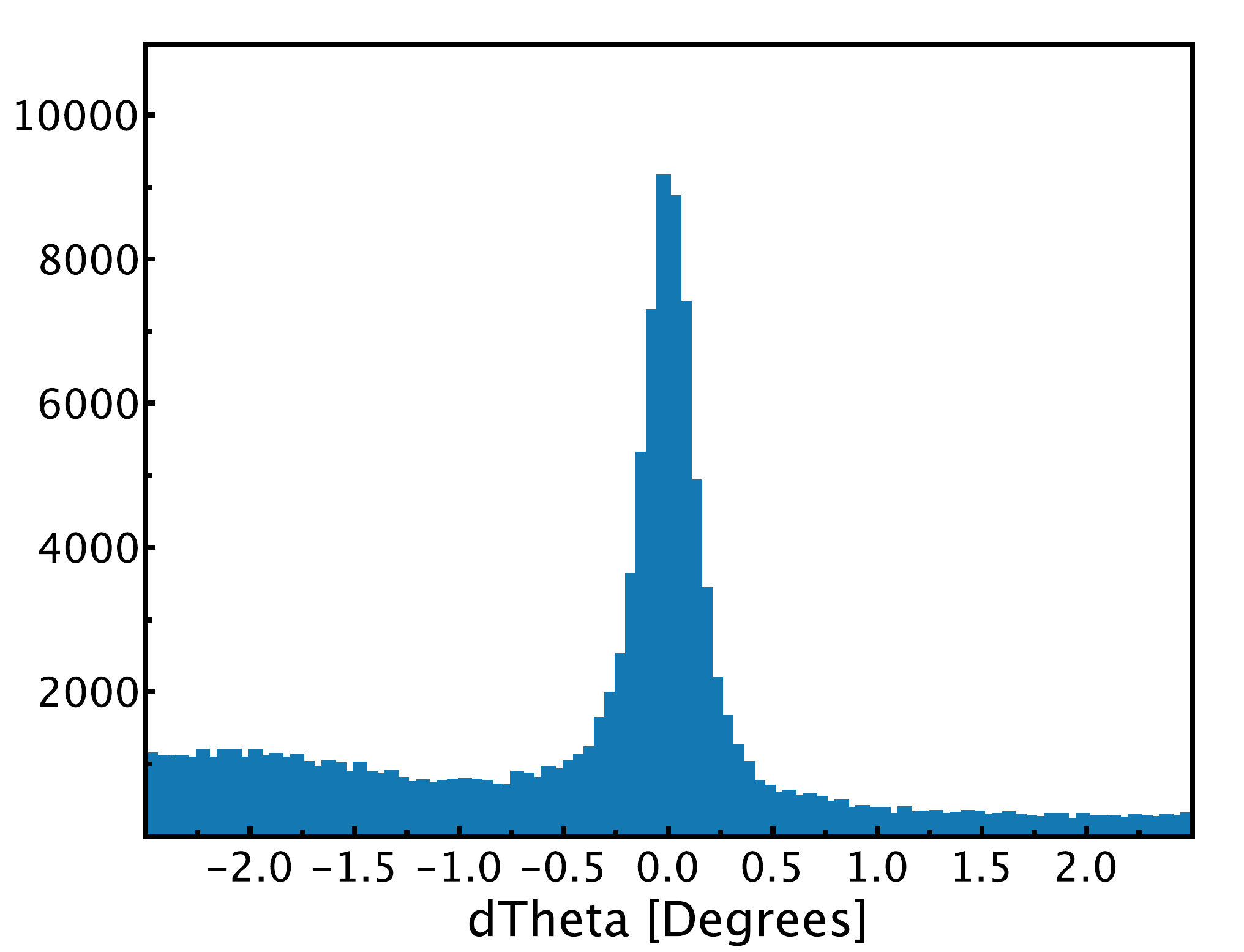}\\
    \caption[The azimuthal and polar angular difference between negative and neutral particles.]{Left: The azimuthal vs polar angular difference between negative and neutral particles. Right: The polar angular difference between negative and neutral particles.}
    \label{fig:dPhi_dTheta}
\end{figure}

\noindent
Another way to identify electrons is to look at negative particles that have a small polar angular distance from neutrals. These correspond to electrons that have radiated a photon ($e^-\to e^-\gamma$) when passing through some material in between the target where the electron was produced and the calorimeter where the electron and photons are detected. Figure~\ref{fig:dPhi_dTheta} shows the azimuthal angular difference ($\Delta\phi$) between negative and neutral particles as a function of the polar angular difference ($\Delta\theta$) . The two peaks close to $\Delta\theta\sim0^o$ are due to photons radiated by electrons. The peak at $\Delta\phi\sim0^o$ is due to photons produced after the drift chambers. The other peak is due to photons produced before the drift chambers and where the trajectory of the electron was curved by the drift chamber magnetic field whereas the trajectory of the photon is unchanged as it does not have an electric charge.\\

\noindent
Figure~\ref{fig:dPhi_dTheta} was made for particles that are not identified as electrons by the offline reconstruction, clearly demonstrating issues with the offline electron identification. To clean up some background under the $\Delta\theta$ peak and ensure only electrons are present, the electrons are required to be associated with at least one hit in the calorimeters. The HTCC is also required to have at least one hit in the same or neighboring sectors, although the hit does not have to be associated to the particle producing the photon. A loose cut on $\Delta\phi$ is placed at $\Delta\phi<|30|^o$. The $\Delta\theta$ distribution is shown in the right histogram of Figure~\ref{fig:dPhi_dTheta} for negative particles in the fiducial region of the PCAL. A cut on $|\Delta\theta|<0.5^o$ is then made to select electrons towards a training sample that does not depend on the offline identification.\\

\noindent
A new training sample is therefore formed from electrons identified by their radiating a photon and from the previous training sample described in Section~\ref{sect_algo} that used the offline identification scheme. The electrons identified with a radiated photon are required to \textbf{not} pass a fiducial cut at 9cm on the calorimeter meaning that they have hits outwith the calorimeter fiducial region. This is done to ensure that both the cases where the offline electron identification is efficient or not are represented in the training sample. Electrons identified with the offline reconstruction represent two thirds of the training sample, whilst electrons identified with a radiated photon represent one third of the training sample. The negative sample uses the offline identification scheme only as improving the electron purity is not a priority and suffers from difficulties in forming training samples from exclusive reactions as explained in the previous section.\\

\begin{figure}[ht!]
    \centering   
    \includegraphics[width=0.7\textwidth]{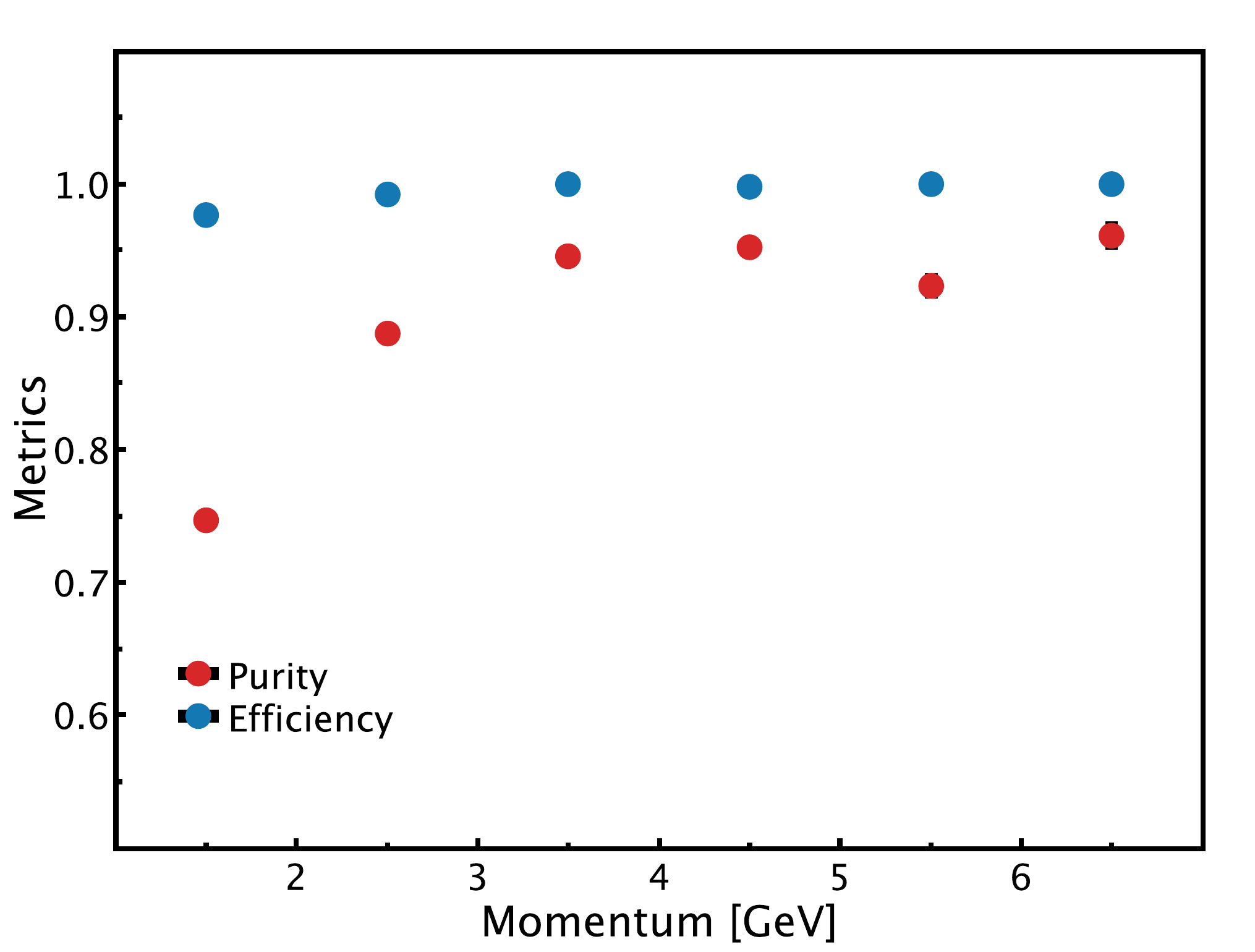}
    \caption[The online electron identification efficiency and purity once retrained to improve the efficiency.]{The online electron identification efficiency and purity relative to the offline electron identification algorithm.}
    \label{fig:mets_p_rad}
\end{figure}

\noindent
Figure~\ref{fig:mets_p_rad} shows the efficiency and purity of the re-trained model relative to the offline electron identification scheme. The electrons are required to pass fiducial cuts placed at 9cm from the edges of the calorimeter. As shown, the online identification is still capable of retaining all electrons as identified by the offline scheme. At high momentum, the purity is decreased by about 5\% to 10\% when compared to the case shown in Figure~\ref{fig:mets_p} where the online identification was trained solely with the offline identification scheme. However, Figure~\ref{fig:MM_n_barchart} showed that the offline identification scheme is inefficient by 7\%. The decrease in purity shown in Figure~\ref{fig:mets_p_rad} will likely be due to electrons that are mis-identified by the offline scheme.\\

\begin{figure}[ht!]
    \centering   
    \includegraphics[width=0.49\textwidth]{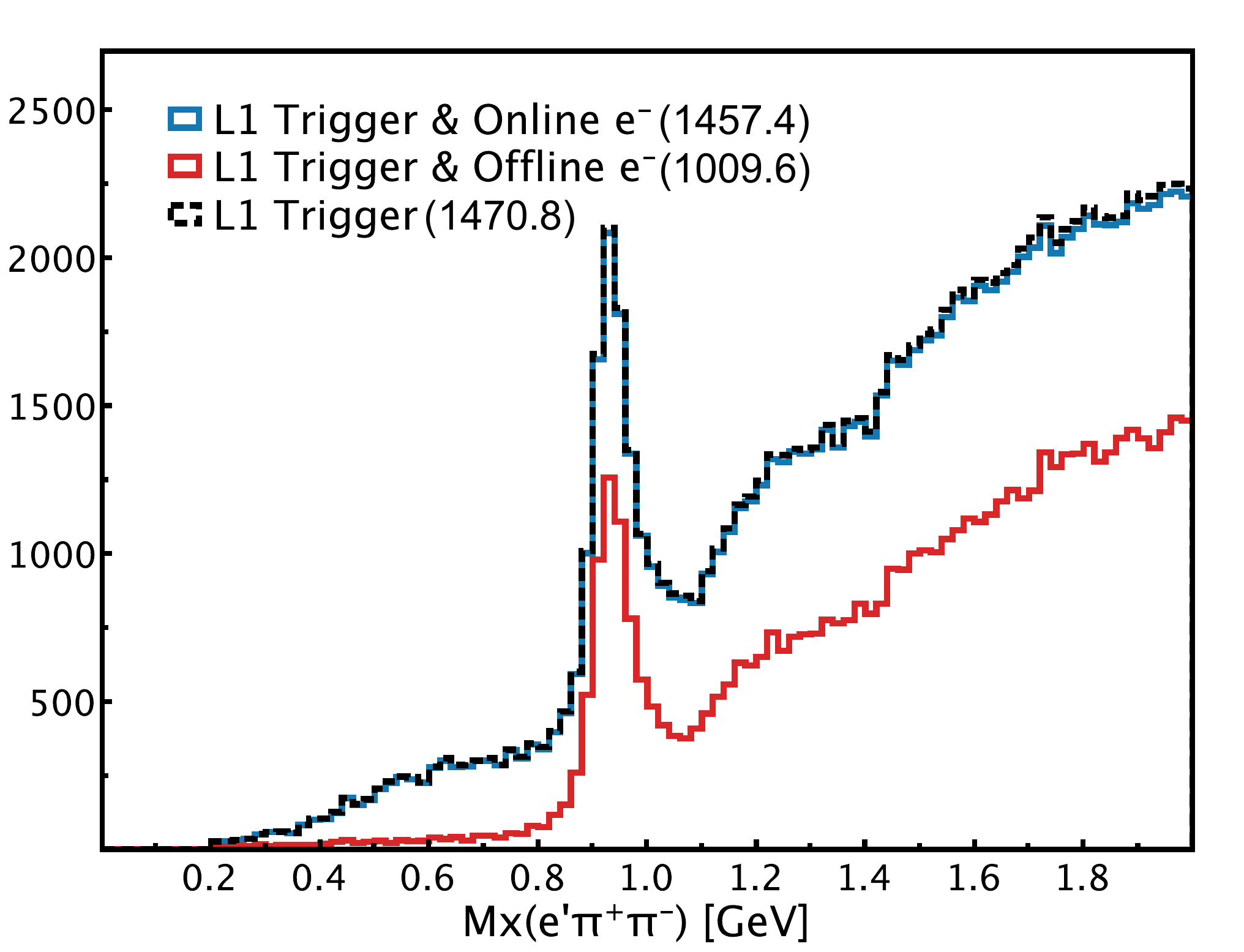}
    \includegraphics[width=0.49\textwidth]{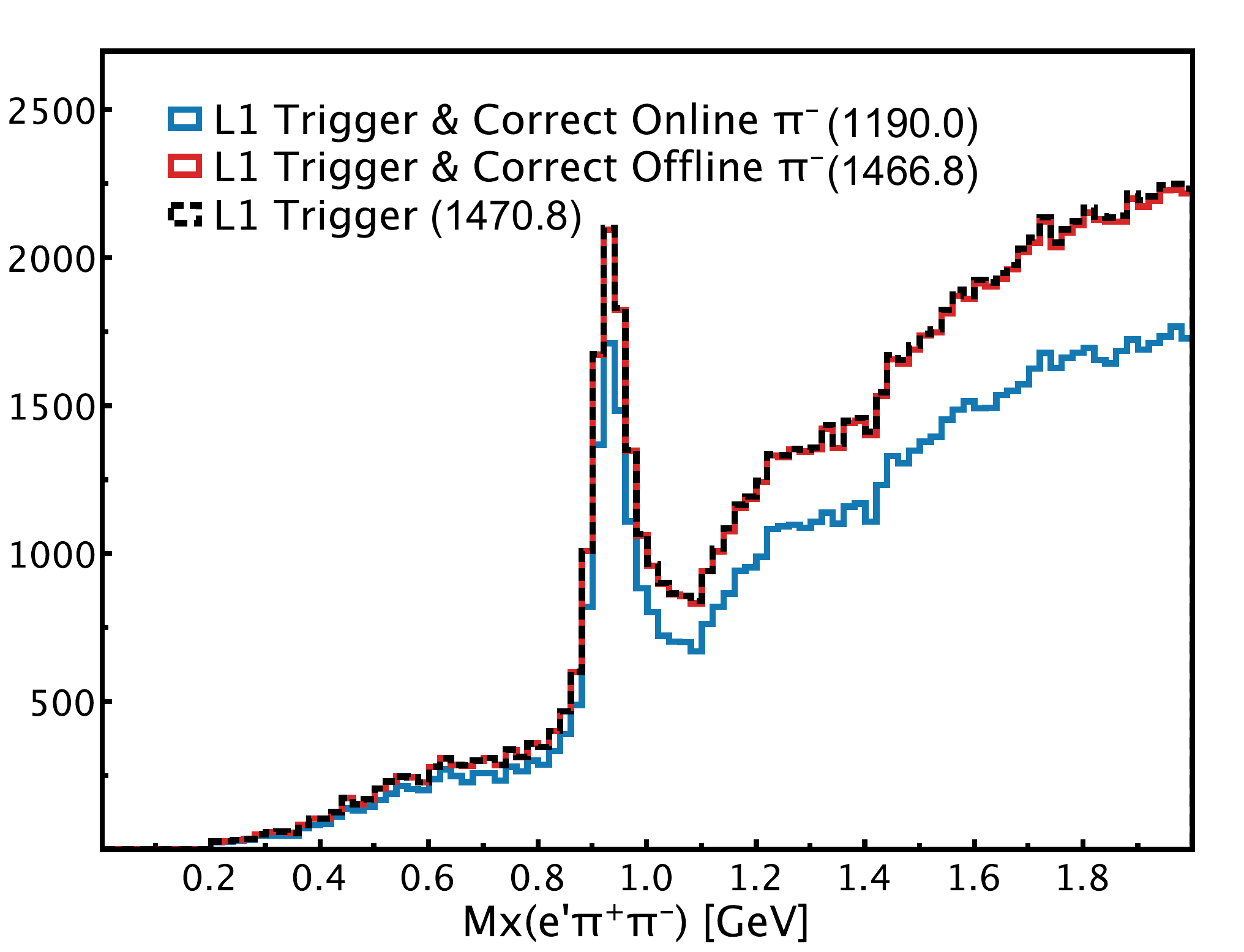}
    \caption[The missing mass of $ep \to e' \pi^+\pi^- X$ after re-training the online scheme to improve its efficiency.]{Left: The missing mass of $ep \to e' \pi^+\pi^- X$ for all events selected by the conventional level 1 electron trigger (black) and events where the electron candidate was identified as an electron by the offline (red) and online (blue) identification schemes. Right: The missing mass of $ep \to e' \pi^+\pi^- X$ for all events selected by the conventional level 1 electron trigger (black) and events where the pion candidate was \textbf{not} identified as an electron by the offline (red) and online (blue) identification schemes. For both left and right, the online identification scheme has been re-trained to improve its efficiency. The missing mass distributions are plotted without cuts on the distance from the edge of the PCAL. The number of electrons estimated from the fit to the proton peak in the missing mass distributions are shown within the legend.}
    \label{fig:MM_n_rad}
\end{figure}

\noindent
Figure~\ref{fig:MM_n_rad} shows the missing mass of $ep \to e' \pi^+\pi^- X$ for all events selected by the conventional level 1 electron trigger and events where the electron candidate was identified as an electron by the offline or online identification schemes once the online scheme has been re-trained to improve its efficiency. Figure~\ref{fig:MM_n_rad} shows the missing mass of $ep \to e' \pi^+\pi^- X$ when no fiducial cuts have been applied to the electrons. The number of electrons estimated by fitting the missing mass distributions are shown in Figure~\ref{fig:MM_n_barchart}. As shown, the online electron identification scheme is able to reach efficiencies above 99\% once re-trained with electrons outwith the calorimeter fiducial region. Figure~\ref{fig:MM_n_rad} also shows the missing mass of $ep \to e' \pi^+\pi^- X$ for all events selected by the conventional level 1 electron trigger and events where the pion candidate was correctly \textbf{not} identified as an electron by the offline and online identification schemes. The purity obtained by fitting the proton peak in the missing mass distributions is of 81\%, which is consistent with Figure~\ref{fig:mets_p_rad}.\\

\noindent
The online identification scheme as described in this section is able to reach efficiencies of about 99\%. The scheme developed here also reaches the high rates shown in Figure~\ref{fig:rates} that are capable of keeping up with the CLAS12 data acquisition rate. As explained above, the purity of the online identification scheme at high momentum is only quoted relative to the offline scheme, despite the knowledge that this is flawed at high momentum. However, the online electron identification found at least one electron in only 60\% of events where the conventional level 1 electron trigger with DC roads found at least one electron. This will lead to a data reduction of 40\% when using the online identification as a second pass filter after the conventional electron trigger. The online identification scheme will provide a useful tool towards the reconstruction of electrons during data taking operations, which will have major implications for the online monitoring and analysis capabilities of CLAS12. It will also be capable of acting as a filter to the conventional trigger, reducing the data rates at CLAS12, and can be used for online event selection which allows for faster post-processing of datasets recorded at CLAS12.\\

\begin{figure}[ht!]
    \centering   
    \includegraphics[width=0.9\textwidth]{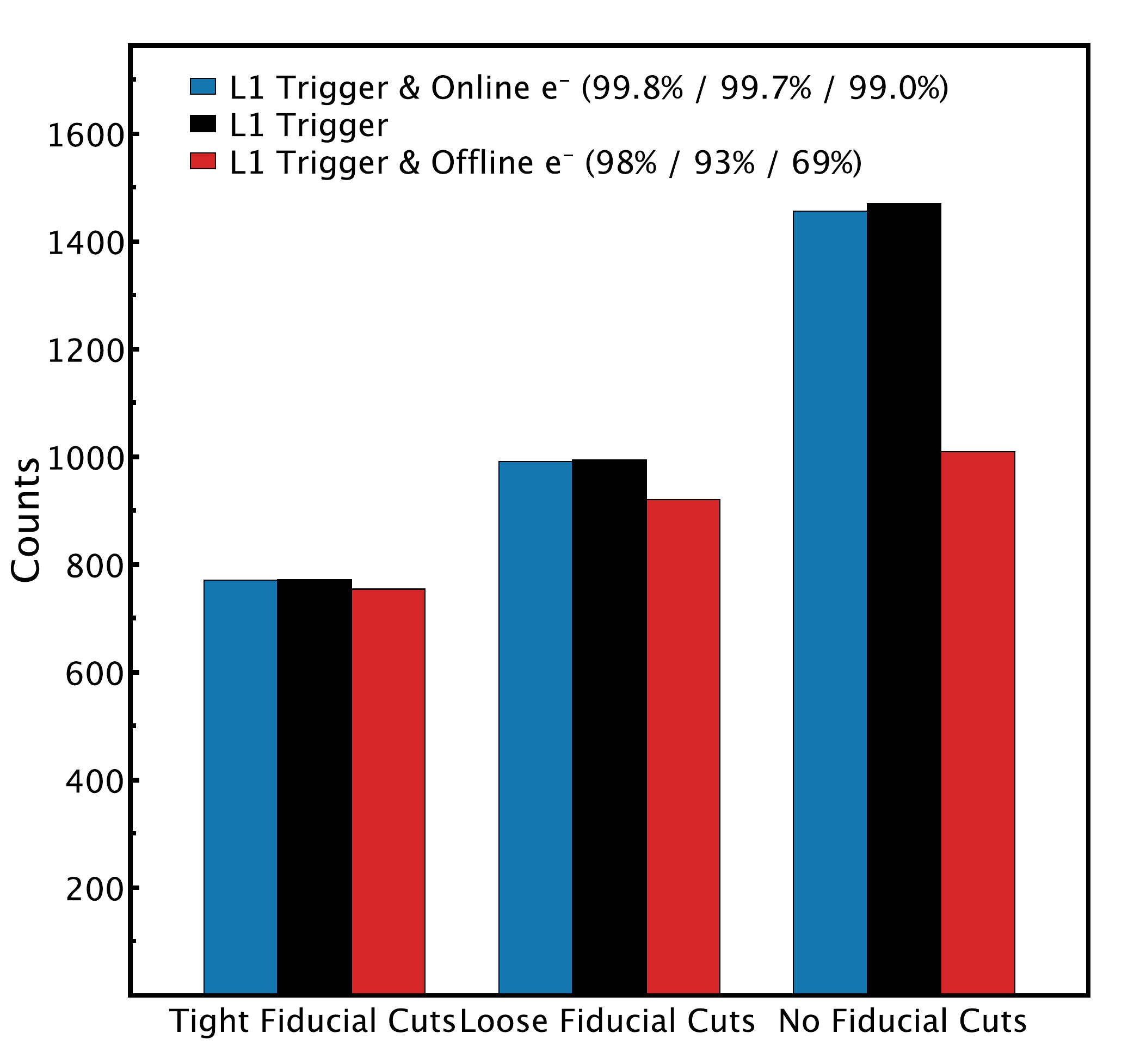}
    \caption[The number of electrons in $ep \to e' \pi^+\pi^- X$ events and identified by the online and offline schemes after re-training the online scheme to improve its efficiency.]{The total number of $ep \to e' \pi^+\pi^- X$ events selected by the conventional level 1 trigger (black) and those where the electron is correctly identified by the online (blue) and offline (red) schemes, for the tight fiducial cuts at 13.5 cm, the looser fiducial cuts at 9 cm and without fiducial cuts. The online identification scheme has been re-trained to improve its efficiency.}
    \label{fig:MM_n_barchart_rad}
\end{figure}

\section{Conclusion and Outlook} \label{sect_ccl}
\noindent
In this article we have demonstrated how full online electron reconstruction could be achieved for the CLAS12 Forward Detector, based on the recent integration of online track reconstruction in data taking operation at CLAS12~\cite{InstaREC}. A first neural network allows to associate a track to hits in the CLAS12 calorimeters. Information at the data acquisition level from the calorimeters and High Threshold Cherenkov Counters associated with the track are then passed to a subsequent neural network that is capable of identifying electrons with an efficiency close to 100\% over the full detector region. Further tests have shown that the online electron identification can be used as a filter to the conventional trigger, allowing for a data reduction of 40\% in standard CLAS12 operating conditions. The scheme proposed in this article would allow to improve data taking operations at CLAS12 by improving its capabilities for real time analyses, monitoring, triggering and online event selection.\\

\noindent
Future work on the online electron identification will present the implementation of the identification scheme within the existing online reconstruction. The general software toolkit to do so exists and is available at Ref.~\cite{CAOS}. Detector material degradation, changes to the run conditions and even changes to the environmental conditions, such as temperature and pressure, may affect the data at the data acquisition level, which in turn would affect the online electron identification performance. The implementation of the online electron identification will require dedicated monitoring to evaluate the performance of the electron identification, for example relative to the conventional electron trigger. In the case where the performance is seen to vary over time, we also envision a continual learning scenario~\cite{Cont} where the online identification networks are retrained based on, for example, significant changes in the number of online electrons relative to the conventional electron trigger. This should in turn stabilise the online electron identification performance at the levels demonstrated in this article.\\

\noindent
Online electron identification capabilities also open an avenue for the identification of charged hadrons. As discussed in Section~\ref{sect_conv}, the electron path to the forward time of flight detector is used to calculate an event start time, based on the time measured at the forward time of flight detector and the assumption that electrons travel at velocities indistinguishable from the speed light at CLAS12 energies and resolutions. Section~\ref{sect_algo} detailed how a neural network could predict with good resolution the path to the time of flight detector and the scintillator component hit by the particle. This information would then allow to find the particle's time at the time of flight detector. For electrons, the time and path would allow to calculate the event start time. This start time is then used to calculate the time of flight for hadrons, which allows to calculate their velocity when combined with their path to the time of flight detector. The velocity then allows to identify charged hadrons.\\

\noindent
Overall, the electron identification scheme proposed in this article allows to efficiently identify electrons, attaining high purities in rejecting other negatively charged particles, and does so at high rates capable of matching the typical CLAS12 data acquisition rates. This will majorly improve the online monitoring and analysis capabilities of CLAS12, which has many potential positive applications. For example, simple analyses such as that of the $ep \to e'\pi^-\pi^+(p)$ reaction presented in Section~\ref{sect_tests} can be used to monitor the quality of the data. Measuring observables such as cross sections and comparing the rates to those estimated from known cross sections could provide an online estimation of the CLAS12 detection efficiency as a function of the running conditions. This could be used to optimise the running conditions to improve the statistics available for analysis. Online analyses can also be used to flag relevant events in the saved datasets, allowing to only post-process targeted events. This could, for example, be greatly beneficial during the calibration of the recorded data as this process requires multiple post processing passes. Online event selection will also be invaluable for experiments employing triggerless data streaming paradigms where online data selection algorithms will only filter events that have a level of noise that would render physics analyses impractical~\cite{ECCE}. As such, efficient event selection procedures implemented during data taking operations will be crucial to handle the large data volume produced by these experiments. Figure~\ref{fig:nonElTrigger} also showed how online electron identification would be useful for non-electron triggering. Finally the online electron scheme is capable of decreasing the recorded data rate by filtering events incorrectly selected by the conventional trigger. This approach to triggering will be beneficial when transitioning to higher luminosity experiments at CLAS12 where the data volume will increase significantly as this work contributes to the reduction in recorded data volumes and offline data processing times. \\

\noindent
The projected Electron-Proton/Ion Collider (ePIC) detector at the future Electron Ion Collider (EIC) envisions running in a triggerless data streaming operating mode~\cite{ECCE}. The solenoidal large intensity device (SoLID) will aim to push the luminosity frontier in electro-production experiments~\cite{SOLID}, whilst the MOLLER Experiment will aim for an high precision measurement of the weak mixing angle using m{\o}ller scattering~\cite{MOLLER}. All three experiments will utilise electron beams and would therefore greatly benefit from in real time electron identification schemes, whether for enhanced online monitoring or improved triggering and online event selection. The approach proposed here, whilst specific to the CLAS12 detector in its use of detector information, is generic enough in design to be replicated and deployed in all three experiments.\\

\section*{Acknowledgements}
\noindent
The authors would like to thank the CLAS Collaboration for providing data used in this body of work.
This material is based upon work supported by the U.S. Department of Energy, Office of Science, Office of Nuclear Physics under contract DE-AC05-06OR23177. The research described in this paper was conducted under the Laboratory Directed Research and Development Program at the Thomas Jefferson National Accelerator Facility for the U.S. Department of Energy.\\



\end{document}